\documentclass[a4paper,11pt]{article}
\pdfoutput=1 

\usepackage{jcappub} 

\usepackage[T1]{fontenc} 
\usepackage{graphicx}
\usepackage{color}
\usepackage{gensymb}

\title{\boldmath Illuminating the dark ages: Cosmic backgrounds from accretion onto primordial black hole dark matter}

\author{G. Hasinger}

\affiliation{European Space Astronomy Centre (ESA/ESAC)\\E-28691
Villanueva de la Cañada, Madrid, Spain}

\emailAdd{guenther.hasinger@esa.int}

\abstract{The recent interpretation of cold dark matter as the sum of contributions of different mass Primordial Black Hole (PBH) families \citep{2019rsta.377..2161G} could explain a number of so far unsolved astrophysical mysteries. Here I assume a realistic $10^{-8}$--$10^{10}$ M$_{\odot}$ PBH mass distribution providing the bulk of the dark matter, consistent with all observational constraints. I estimate the contribution of baryon accretion onto this PBH population to various cosmic background radiations, concentrating first on the cross-correlation signal between the Cosmic X--ray and the Cosmic infrared background fluctuations discovered in deep {\em Chandra} and {\em Spitzer} surveys. I assume Bondi capture and advection dominated disk accretion with reasonable parameters like baryon density and effective relative velocity between baryons and PBH, as well as appropriate accretion and radiation efficiencies, and integrate these over the PBH mass spectrum and cosmic time. 
The prediction of the PBH contribution to the X--ray background is indeed consistent with the residual X--ray background signal and the X--ray/infrared fluctuations. The predicted flux peaks at redshifts z$\approx$17--30, consistent with other constraints requiring the signal to come from such high redshifts. The PBH contribution to the 2--5 $\mu$m cosmic infrared background fluctuations is only about 1\%, so that these likely come from star formation processes in regions associated with the PBH.

\vskip 0.1 truecm
I discuss a number of other phenomena, which could be significantly affected by the PBH accretion. Magnetic fields are an essential ingredient in the Bondi capture process, and I argue that the PBH can play an important role in amplifying magnetic seed fields in the early universe and maintaining them until the galactic dynamo processes set in. Next I study the contribution of the assumed PBH population to the re-ionization history of the universe and find that they do not conflict with the stringent ionization limits set by the most recent {\em Planck} measurements. X--ray heating from the PBH population can provide a contribution to the entropy floor observed in groups of galaxies. The tantalizing redshifted 21-cm absorption line feature observed by {\em EDGES} could well be connected to the radio emission contributed by PBH to the cosmic background radiation. Finally, the number of intermediate-mass black holes and the diffuse X--ray emission in the Galactic Center region are not violated by the assumed PBH dark matter, on the contrary, some of the discrete sources resolved in the deepest {\em Chandra} observations of the Galactic Ridge could indeed be accreting PBH.}

\begin{document}
\maketitle
\flushbottom

\section{Introduction}
\label{sec:intro}

Recent years saw a revival of the idea originally put forward by S. Hawking \cite{1971MNRAS.152...75H}, that Primordial Black Holes (PBH) could make up the so far elusive Dark Matter. LIGOs first detection of gravitational waves from merging binary black holes of approximately equal masses in the range 10--30 M$_{\odot}$ \citep{2016PhRvX...6d1015A,2019ApJ...882L..24A} led to the suggestion that these could be a signature of dark matter stellar mass PBH \citep{2016PhRvL.116t1301B,2016ApJ...823L..25K,2017PDU....15..142C} in a mass window not yet excluded by other astrophysical constraints. A recent review about the rich literature constraining the possible contributions of PBH to the dark matter is e.g. given in \cite{2019PhRvD.100d3540M}. 

\vskip 0.1 truecm
In a recently published theoretical prediction \cite{2019rsta.377..2161G,2019arXiv190411482G} PBH are created in the QCD phase transitions (around 100 MeV) of different particle families freezing out of the primordial Quark-gluon plasma within the first two seconds after the inflationary phase. When W$^{+/-}$, Z bosons, baryons, pions are created, and e$^+$e$^-$ pairs annihilate, they leave an imprint in form of a significant reduction of the sound speed at the corresponding phase transitions, and allow regions of high curvature to collapse and form PBH \citep[see also][]{2018JCAP...08..041B}. The typical mass scale of these PBH is defined by the size of the horizon at the time of the corresponding phase transition. In this model four distinct populations of PBH in a wide mass range are formed: planetary mass black holes at the W$^{+/-}$, Z transition, PBH of around the Chandrasekhar mass when the baryons (protons and neutrons) are formed from 3 quarks, PBH of masses of order 30 M$_{\odot}$ (these correspond to the LIGO black holes), when pions are formed from two quarks, and finally supermassive black holes (SMBH) at the e$^+$e$^-$ annihilation \citep[see also][]{2017JPhCS.840a2032G}. Another remarkable aspect of this theory is, that the gravitational energy released at at the PBH collapse locally reheats regions (hot spots) around the black holes to the electroweak transition scale (around 100 GeV), where chiral sphaleron selection effects can introduce the matter/antimatter asymmetry. The PBH in this picture would therefore also be responsible for the baryogenesis and fix the ratio of dark matter to baryons. Clustering of the PBH in a very wide mass distribution could alleviate some of the more stringent observational constraints on the allowed contribution of PBH to the dark matter \citep{2017PDU....15..142C,2019EPJC...79..246B}. The interpretation of cold dark matter as the sum of contributions of different mass PBH families could explain a number of so far unsolved mysteries, like e.g. the massive seed black holes required to create the supermassive black holes in the earliest QSOs \citep{2007ApJ...665..187L}, the ubiquitous massive LIGO/VIRGO massive binary black holes \citep[e.g.][]{2016ApJ...823L..25K}, or even the putative "Planet X" PBH in our own Solar System \citep{2019arXiv190911090S}.

\vskip 0.1 truecm
The most abundant family of PBH should be around the Chandrasekhar mass (1.4 M$_{\odot}$). This prediction may already have been vindicated by the recent OGLE/GAIA discovery of a sizeable population of putative black holes in the mass range 1--10 M$_{\odot}$ \citep{2019arXiv190407789W}. The microlensing survey OGLE has detected $\sim$60 long-duration microlensing events. About 20 of these have GAIA DR2 parallax distances of a few kpc, which break the microlensing mass–distance degeneracy and allow the determination of masses in the few solar mass range, implying that these objects are most likely black holes, since stars at those distances would be directly visible by OGLE. 

\vskip 0.1 truecm
Important fingerprints of a population of PBH may be hidden in the Cosmic infrared and X--ray background radiation (see \cite{2018RvMP...90b5006K} for a comprehensive review). Indeed, \cite{2016ApJ...823L..25K} argues, that the near-infrared Cosmic background (CIB) anisotropies detected in deep {\em Spitzer} \cite{2005Natur.438...45K,2007ApJ...654L...5K,2010ApJS..186...10A,2012ApJ...753...63K} and {\em Akari} \cite{2011ApJ...742..124M} images, which cannot be accounted for by known galaxy populations \cite{2012ApJ...752..113H}, could be connected to PBH. Similar fluctuations were discovered in the Cosmic X--ray background (CXB) observed in a deep {\em{Chandra}} survey, which are correlated with the CIB anisotropies in the same field \cite{2013ApJ...769...68C}.  Later studies of wider/deeper fields covered by both {\em Chandra} and {\em Spitzer} \citep{2017ApJ...847L..11C,2018ApJ...864..141L,2019ApJ...883...64L} have substantially improved the detection significance of the observed signal. The X--ray fluctuations contribute about 20\% to the CIB signal, indicating that black hole accretion should be responsible for such highly efficient X--ray emission. Similar studies of deep fields observed with the {\em{Hubble}} Space Telescope in the optical range do not show such a cross-correlation signal down to m$_{AB}$$\sim$28 \citep[see][]{2018RvMP...90b5006K}. The angular scale of the fluctuation power spectra of the CIB and CXB reach values >1000", much larger than expected for the known galaxy populations \citep{2014ApJ...785...38H}. All of these findings can be understood, if the fluctuation signal comes from a high-redshift (z$\gtrsim$12) population of black holes. The spectral shape of the CXB fluctuations determined from a combination of the deepest/widest fields \cite{2019ApJ...883...64L} can be fit either with a very high redshift population of obscured black holes, or with completely unobscured black hole accretion. Original models \citep{2013MNRAS.433.1556Y} invoked highly obscured Direct Collapse Black Holes formed in metal-free halos at z>12 to explain the observed CIB and CXB signal. However,  accreting massive black holes have recently been firmly ruled out as the source of these fluctuations \cite{2019MNRAS.489.1006R}, because they would require an unfeasible amount of black hole accretion at z>6, locking up a larger amount of mass in massive black holes at high redshift, than the known black hole mass function at z=0. These authors also ruled out local diffuse emission as the source of the X--ray fluctuations. The CXB has been largely resolved into discrete sources in deep X--ray images, either directly \citep[see][]{2005ARA&A..43..827B,2006ApJ...645...95H}, or by crosscorrelating with the deepest {\em Hubble} galaxy catalogues \citep{2007ApJ...661L.117H,2012ApJ...748...50C}. However, \cite{2007ApJ...661L.117H} show that some marginally significant diffuse CXB still remains after accounting for all discrete contributions. This is consistent with the independent determination of \cite{2017ApJ...837...19C}. The residual unresolved flux is about 3 times larger than the X-ray flux associated with the above CXB/CIB fluctuations. 

\vskip 0.1 truecm
Given the difficulties in explaining the CIB/CXB correlation with known classes of sources, and motivated by the notion that the dark matter could be dominated by an extended mass distribution of PBH, I constructed a toy model to explore the potential contribution to the cosmic backgrounds by the accretion of baryons throughout cosmic history onto such a population of early black holes. Assuming a combination of Bondi-Hoyle-Lyttleton quasi-spherical capture at large distances from the PBH, and advection-dominated disk accretion flows (ADAF) in the vicinity of the central object, I can explain the observed residual CXB flux and the CXB/CIB crosscorrelation with minimal tuning of the input parameters, and find a maximum contribution to the extragalactic background light in the redshift range 15<z<30. I further estimate that this accretion onto PBH can produce enough flux to significantly contribute to the pre-ionization of the intergalactic medium with UV photons by a redshift z$\gtrsim$15 and to the pre-heating of the baryons with X--ray photons, observed as an "entropy floor" in the X--ray emission of galaxy groups.

\vskip 0.1 truecm
In section 2 the assumed PBH mass distribution is introduced and contrasted with recent observational limits on the PBH contribution to the dark matter. The basic ingredients of the toy model for the accretion onto PBH are presented in section 3. The assumed radiation mechanism and efficiency is discussed in section 4. The contribution of the PBH emission to the different bands is compared with the observational constraints in section 5. Other potential diagnostics of this putative dark matter black hole population are discussed in section 6, and conclusions are presented in section 7. Throughout this work a $\Lambda$CDM cosmology with $\Omega_M$=0.315, $\Omega_\Lambda$=0.685, and H$_0$=67.4 km s$^{-1}$  Mpc$^{-1}$ \citep{2018arXiv180706209P} is used. These parameters define the baryon density $\Omega_{bar}$=0.049, the dark matter density $\Omega_{DM}$=0.264, and the critical mass density of the universe $\rho_{crit}$=1.26$\times10^{20} M_\odot~ {\rm Gpc}^{-3}$. All logarithms in this paper are taken to the base 10.

\section{The assumed PBH mass distribution}
\label{sec:PBH}

The theoretical predictions in \cite{2019rsta.377..2161G,2019arXiv190411482G,2017JPhCS.840a2032G,2019arXiv190608217C} yield a broad distribution of PBH masses with a number of peaks corresponding to the particle families freezing out from the Big Bang. Depending on the spectral index $n_s$ of the primordial curvature fluctuation power spectrum, the PBH mass distribution has a different overall slope. \cite{2019arXiv190608217C} find consistency of these predictions with a number of recent observational limits on the PBH contribution to the dark matter, but there is a tension of their models with the Cosmic Microwave Background (CMB) constraints from accretion at large PBH masses \citep{2017PhRvD..96h3524P,2017PhRvD..95d3534A}. Recent limits from gravitational lensing of type Ia supernovae on a maximum contribution of stellar-mass compact objects to the dark matter of around 35\% \citep{2018PhRvL.121n1101Z}, and from the LIGO OI gravitational wave merger rate of black holes in the mass range 10--300 M$_{\odot}$ \cite{2017PhRvD..96l3523A} are also in tension with these models. An additional important constraint comes from a comparison of the predicted PBH fraction with the measured local mass function of supermassive black holes (SMBH) in the centers of nearby galaxies. Integrating the local SMBH mass function of \cite{2013CQGra..30x4001S} (see figure \ref{fig:fPBH}) in the range $10^{6}$--$10^{10}$ M$_{\odot}$ yields a local SMBH mass density of $\rho_{SMBH}$=6.3$\times$10$^5$ M$_{\odot}$ Mpc$^{-3}$, corresponding to a dark matter fraction of f$_{SMBH}$=1.89$\times$10$^{-5}$, which is about a factor of 10--100 lower than the f$_{PBH}$ predictions in \cite{2019rsta.377..2161G,2019arXiv190608217C}.

\begin{figure*}[htbp]
    \begin{center}
    \includegraphics[width=0.6\textwidth]{figures/f_PBH.jpg}
    \end{center}
    \caption{The PBH mass spectrum (thick red line) assumed for this work (Garc{\'\i}a-Bellido, 2020, priv. comm.), compared to a number of observational constraints. Microlensing limits from SNe \citep{2018PhRvL.121n1101Z}, EROS \citep{2007A&A...469..387T}, and the Subaru M31 survey \citep{2019NatAs...3..524N} are shown as solid, dashed and dotted green lines, respectively. LIGO limits from gravitational merger event rates are shown as blue solid line for subsolar masses \cite{2019PhRvL.123p1102A}, and as blue dashed line for 10-300 M$_{\odot}$ \cite{2017PhRvD..96l3523A}. The CMB accretion limits from \cite{2017PhRvD..96h3524P} are shown as orange dashed line. Multiwavelength limits from the Galactic Center \cite{2019JCAP...06..026M} are shown in magenta for X-ray (solid) and radio (dashed) observations. Finally, the local SMBH mass function \citep{2013CQGra..30x4001S} is shown as black line at 10$^{6-10}$ M$_{\odot}$.}
    \label{fig:fPBH}
\end{figure*}

\vskip 0.1 truecm
For these reasons, Garc{\'\i}a-Bellido et al. (2020 in prep.) are revising their model parameters in order to predict a steeper PBH mass function at large M$_{PBH}$ and shared one of their new models, shown as red curve in figure \ref{fig:fPBH}. Here a value of n$_s$=0.987 is assumed for the spectral index of the primordial fluctuation power spectrum, as well as a running curvature of dn$_s$=$-$0.0006. The integral of this PBH distribution over the whole mass range yields f$_{PBH}$=1. On the other hand, the distribution yields only $\sim$40\% of the dark matter in the peak mass range [0.1,10] M$_{\odot}$, and is thus fully consistent with the microlensing constraints in figure \ref{fig:fPBH}. In the mass range of the LIGO black hole binaries it predicts just the right amount of dark matter to explain the gravitational wave merger rates, and in the SMBH range it is consistent with the local black hole mass function (taking into account the accretion onto supermassive PBH over cosmic time producing the bulk of the X-ray background \citep{2015A&A...574L..10C}). Apart from small sections, the new PBH mass function is thus fully consistent with the most recent observational constraints.

\section{Baryon accretion onto the PBH}
\label{sec:toy}

In the following I use the PBH mass spectrum presented in section \ref{sec:PBH} to calculate the accretion of baryons onto PBH over cosmic time, and to predict the electromagnetic emission from this process. As we will see, for most of the cosmic history these black holes move at supersonic speeds among the baryons and will therefore undergo Bondi-Hoyle-Lyttleton quasi-spherical capture \citep{1939PCPS...35..405H,1944MNRAS.104..273B,1952MNRAS.112..195B,2004NewAR..48..843E}. In the Bondi-Hoyle picture of a black hole moving supersonically through a homogeneous gas, the capture happens in the wake of the moving object. Behind the object, material moves in from a wide cone, and needs to lose angular momentum before it can fall towards the black hole. The gas is in principle collisionless, so that only the magnetic field trapped in the plasma allows particles to lose angular momentum and start to behave like a fluid. This gas forms the accretion flow, in which it is adiabatically heated. The accreting gas is ionized and embedded in the magnetic field. Any plasma drawn in by the gravitational field will carry along the magnetic field. Shvartsman \cite{1971SvA....15..377S} argues that in the black hole tail, where the matter flow stops, the gravitational and magnetic energy densities become nearly equal. This equipartition is preserved in the infalling flow and thus the magnetic field grows towards the black hole. Like the heat has to be ultimately radiated away, the magnetic field needs a way to dissipate energy on its way inward. \cite{2005A&A...440..223B} discuss that the most likely dissipation mechanism for the magnetic field is reconnection of field lines in narrow current sheets, similar to the processes we observe in solar flares and active galactic nuclei. Magnetic reconnection causes the acceleration and non-thermal heating of a small fraction of the infalling electrons. In parallel, decoupled magnetic field lines can carry some of the amplified magnetic field outward and eject plasma \citep{2005A&A...440..223B}. 

\vskip 0.1 truecm
An important question is, whether the accretion flow is spherically symmetric close to the black hole, or whether an accretion disk is formed. Originally most researchers assumed spherical accretion for PBH \citep[e.g.][]{2007ApJ...662...53R,2008ApJ...680..829R,2017PhRvD..95d3534A}. However, \cite{2017PhRvD..96h3524P} argues, that the accreted angular momentum is large enough, that an accretion disk is formed, at least close to the black hole. According to these authors, the granularity of the PBH distribution and the formation of PBH binaries at the scale of the Bondi radius will imprint density and velocity gradients into the relative distribution of baryons and PBH, such that ultimately an accretion disk and an advection-dominated accretion flow (ADAF) will form \cite{2014ARA&A..52..529Y}. The formation of an ADAF disk significantly reduces the accretion rate and the radiative efficiency \citep{2012MNRAS.427.1580X}, compared to spherical accretion.
But to first order the Bondi-Hoyle-Lyttleton mechanism can be used to estimate the accretion rate $\dot M$ onto the PBH \cite{2017PhRvD..96h3524P,2019PhRvD.100d3540M}.

\vskip 0.1 truecm
Bondi \cite{1952MNRAS.112..195B} discusses two different approximations to the spherical gas accretion problem, (i) the velocity-limited case, where the motion of the accreting object through the gas is dominant and an accretion column is formed in the wake of the moving object, and (ii) the temperature-limited case, where the sound speed of the gas is dominant and a spherical accretion flow forms. In the velocity-limited case (i) the mass accretion rate is given as 
\begin{equation}
    \dot M = 2.5 \pi \rho  (G M)^2 v_{rel}^{-3},
\end{equation} 
where $\rho$ is the gas density, $M$ is the PBH mass, and $v_{rel}$ is the relative velocity between object and gas. In the temperature-limited case (ii) with negligible relative velocity, the thermal velocity of the gas particles is dominant and the corresponding accretion rate is given by
\begin{equation}
    \dot M = 2.5 \pi \rho  (G M)^2 c_s^{-3}, 
\end{equation}
where $c_s$ is the sound speed. For intermediate cases, \cite{1952MNRAS.112..195B} introduces an effective velocity 
\begin{equation}
    v_{eff} = \sqrt{v_{rel}^2+c_s^2}
\end{equation}
and the corresponding mass accretion rates becomes
\begin{equation}
    \dot M = 2\lambda \pi \rho  (G M)^2 v_{eff}^{-3},
\end{equation}
 where the so called accretion eigenvalue $\lambda$ is is a fudge factor of order unity, dependent on non-gravitational aspects of the problem, like e.g. the gas equation of state or outflows from feedback effects. Different authors have discussed this parameter for the particular application of gas accretion onto PBH in the early universe. \cite{2007ApJ...662...53R} find values of $1.12>\lambda>10^{-3}$, depending e.g. on the PBH mass. For masses of order 1 M$_\odot$ they find $\lambda=1.12$.  \cite{2016PhRvL.116t1301B} discriminate between isothermal and adiabatic gas with accretion eigenvalues of $\lambda$=1.12, and 0.12, respectively. In this paper I assume an eigenvalue $\lambda$=0.05. The motivation for this choice is discussed in section 4, while section 5 and 6 show that this choice fits the observational constraints quite well.

\begin{figure*}[htbp]
    \includegraphics[width=0.49\textwidth]{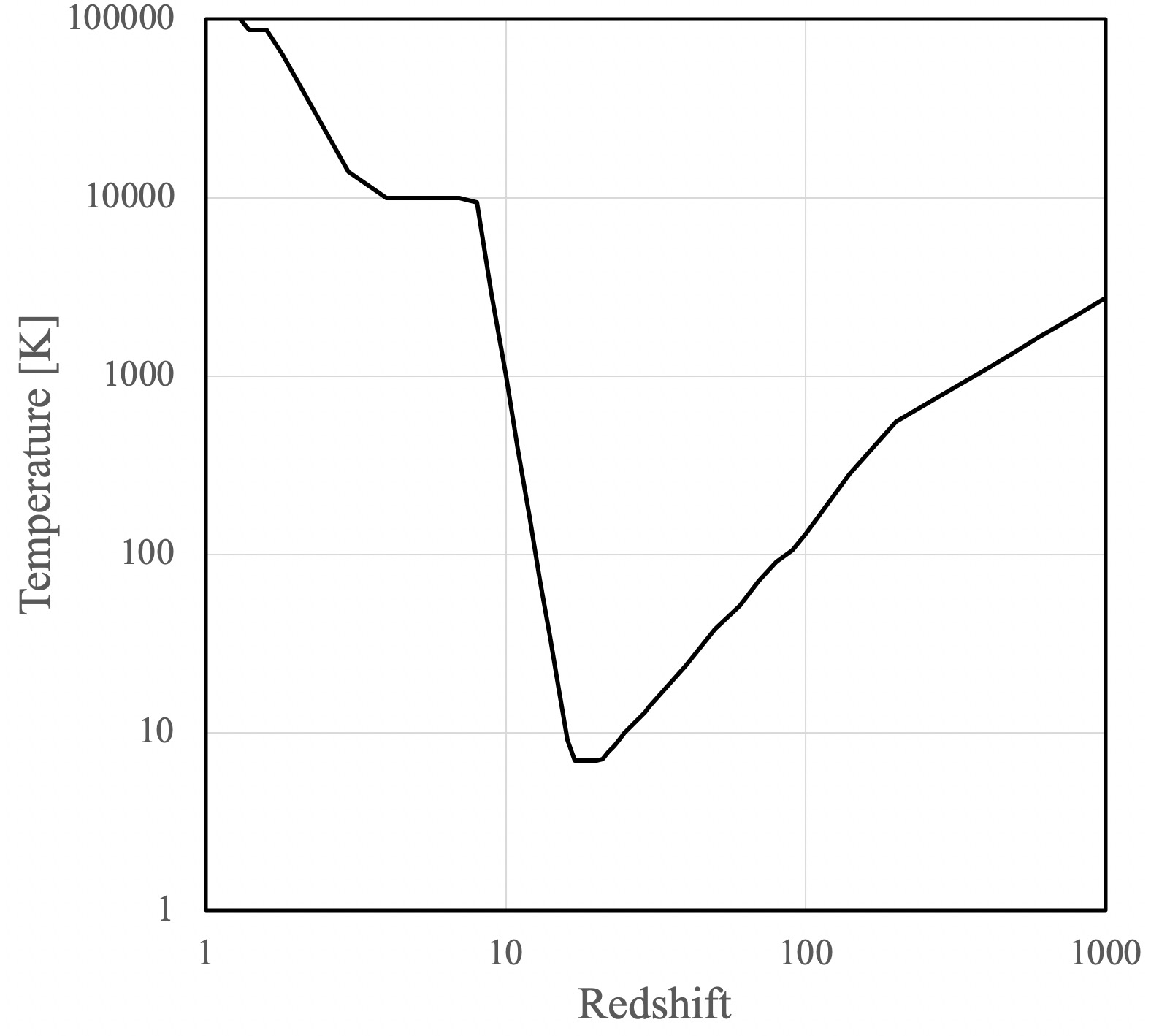}
    \includegraphics[width=0.49\textwidth]{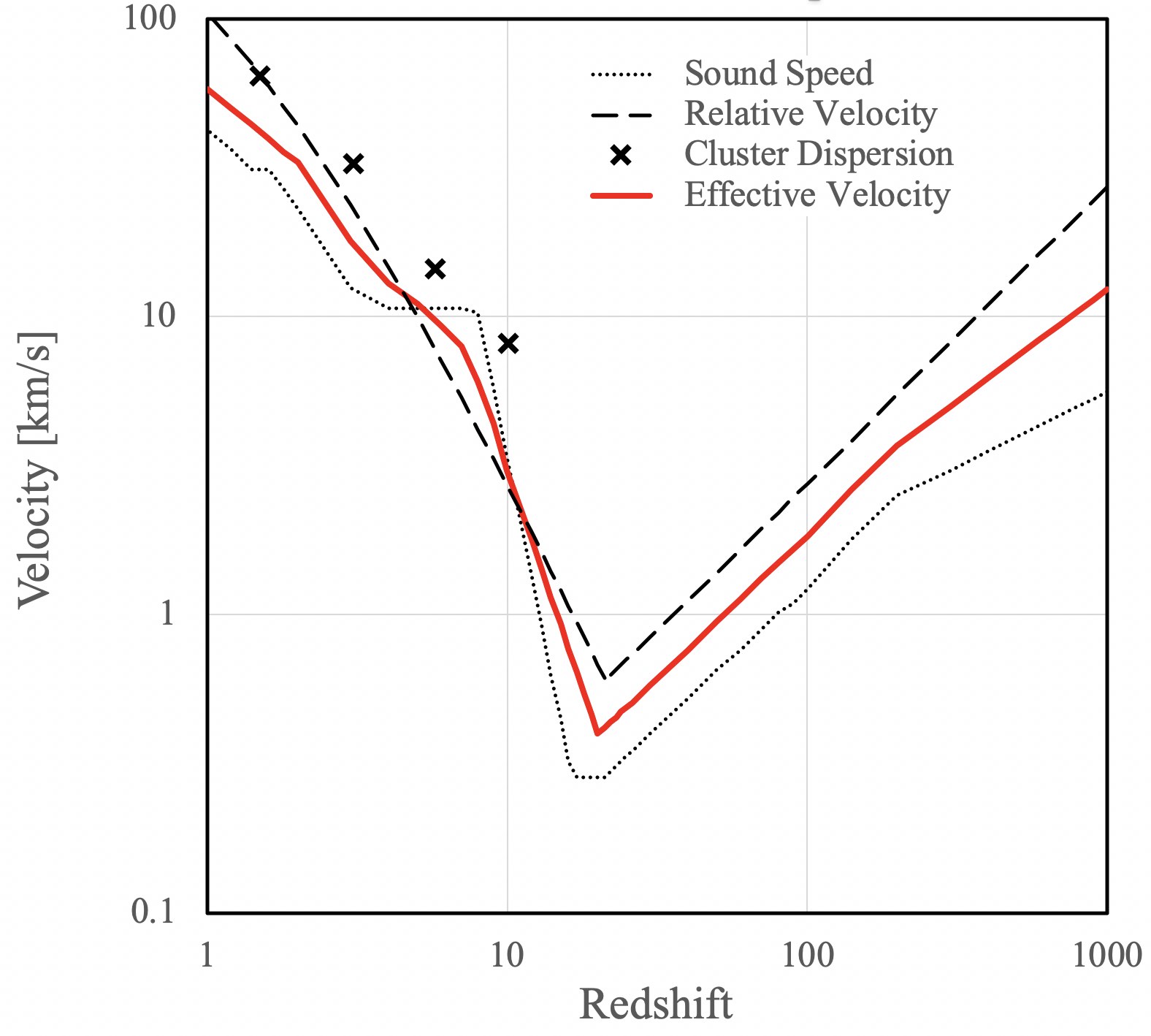}
    \caption{Left: Baryon temperature as a function of redshift. Right: Mean relative velocity $\langle v_{rel} \rangle $ between dark matter and baryons, sound speed $c_s$ and the effective velocity $v_{eff}$ (eq. 3.8) as a function of redshift.}
    \label{fig:TempVel}
\end{figure*}

\vskip 0.1 truecm
Let us first look at the thermal history and thus the sound speed of the gas over cosmic history. A nice summary is given in figure 15 of \cite{2013ASSL..396...45Z}.  Despite having decoupled from the CMB at z$\approx$1089, the gas temperature continues to follow the temperature evolution T$\propto$(1+z) of the background photons due to Compton scattering off residual ionized electrons from the recombination era. Below redshifts z$\approx$200 the residual ionization in the gas is low enough, that it decouples from the background radiation and cools adiabatically following the relation T$\propto$(1+z)$^2$. When the first objects form and reionization starts around z$\lesssim$20, the gas is heated up to temperatures $\sim$10$^4$ K. The details of re-ionization are still uncertain and will be discussed below. I have deliberately chosen a redshift of z$\approx$20 for re-ionization to become dominant, with full ionization occurring around z$\approx$7. Finally, at z<3, when the bulk of the cosmic baryons are falling into increasingly larger dark matter halos and become virialized, they are further heated up to form the warm/hot intergalactic medium at temperatures $10^{5-7}$K \citep{1999ApJ...514....1C}. Using figure 2b in this paper I estimate average temperatures for the IGM of 5$\times10^4$, 1.5$\times10^5$, and 8$\times10^5$ K at z=2, 1, 0, respectively. The baryon temperature as a function of redshift assumed in this work is shown in figure \ref{fig:TempVel} (left). The sound speed of the gas is given by 
\begin{equation}
    c_s=\sqrt{\frac {\gamma k T} {\mu m_H}},
\end{equation} where $\gamma$=5/3 for an ideal monoatomic gas, $\mu$=1.22 is the mean molecular weight including a helium mass fraction of 0.24, $m_H$ is the mass of the hydrogen atom, and $T$ is the temperature of the baryons as a function of cosmic history discussed above \citep{2010PhRvD..82h3520T}. The sound speed as a function of redshift is the dotted curve shown in figure \ref{fig:TempVel} (right).

\vskip 0.1 truecm
I now discuss the relative velocity $v_{rel}$ between the dark matter PBH and the baryons throughout cosmic history.
In the radiation-dominated phase of the universe at z>1089, the dark matter is already hierarchically clustering under the influence of its own gravity. The sound speed of the photon-baryon fluid is very high, of order one third of the velocity of light, and thus the normal matter undergoes baryonic acoustic oscillations \citep{1970Ap&SS...7....3S,1970ApJ...162..815P}. This leads to a spatial separation between baryons and dark matter and thus to a Gaussian distribution of relative velocities with an average around $\langle v_{rel} \rangle$$\approx$30 km/s \citep[see][]{2010PhRvD..82h3520T,2014IJMPD..2330017F}.  At z$\approx$1089, when electrons and protons combine and the universe becomes transparent, the sound speed of the gas dramatically drops to $\sim$6 km/s. The dark matter PBH kinematically decouple from the baryons and their relative velocities become highly supersonic. In the linear growth phase of the universe, at scales larger than the gas Jeans-length, the dark matter and the baryons fall in the same gravitational potentials of the cosmic web and thus their relative velocity decreases with the cosmic expansion: 
\begin{equation}
    \langle v_{rel} \rangle_{linear} \approx 30~{\frac {1+z} {1000}}~{\rm km~s}^{-1}.
\end{equation}

\vskip 0.1 truecm
 This relation is shown as the right branch of the dashed line in figure \ref{fig:TempVel} (right), above redshifts $z\gtrsim20$. 
 From this figure it becomes apparent, that between recombination and re-ionization the PBH move with highly supersonic, but decreasing velocities through the gas, due to the decaying sound waves. As we will see below, in this branch of the velocity curve the contribution of PBH to the cosmic backgrounds has a maximum. At lower redshifts, at scales smaller than the gas Jeans-length, the hierarchical clustering becomes non-linear and baryons falling into growing dark matter halos are virialized. As a consequence, the velocity dispersion between dark matter and gas increases again towards lower redshifts, scaling as $M_{Halo}^{1/3}$, where $M_{Halo}$ is the mass of the dark matter halo becoming non-linear. I used two different methods to estimate the average virial velocity for redshifts z$\lesssim$20. First, the Millenium Simulation run described in \cite{2005Natur.435..629S} gives the mass function of dark matter halos with halo masses $M_{Halo}$>$10^{10} M_\odot$ for five different redshifts between z=10 and z=0. I extrapolated these simulated mass functions to lower masses ($M_{Halo}$>10$^3 M_\odot$) using the empirical universal halo mass function shape found through simulations by \cite{2013MNRAS.433.1230W}.  For every mass bin I determined the virial velocity according to the calibration of the velocity dispersion as a function of halo mass described in \cite{2013MNRAS.430.2638M}, and then calculated the average for each epoch. These virial velocities are shown as crosses in figure \ref{fig:TempVel} (right). The extrapolation to halo masses as small as $M_{Halo}>10^3 M _\odot$ is rather uncertain, both for the mass function and the velocity dispersion, because the cosmological simulations do not have a fine enough mass resolution at this scale. Also, the velocity dispersion relevant for Bondi capture onto PBH is determined by the smallest mass scales becoming non-linear at any redshift. A second possibility to calculate the relative velocities in the non-linear phase is therefore to determine the velocity dispersion directly from the dark matter power spectrum and integrate this over the smallest non-linear scales. This calculation has been performed by M. Bartelmann (2020, priv. comm.), adopting the normalization of the primordial power spectrum of $\sigma_8$=0.8. The relative velocity in the non-linear regime can be approximated by 
 \begin{equation}
     \langle v_{rel} \rangle _{nonlinear} \approx 620~(1+z)^{-2.3}~{\rm km~s}^{-1},
 \end{equation}
 and is shown as the left branch ($z\lesssim20$) of the dashed line in figure \ref{fig:TempVel} (right). At z=2 the cluster velocity dispersion agrees with this estimate, but it systematically overestimates the small-scale velocity dispersion at larger redshifts.

\vskip 0.1 truecm
Since we are interested in the total contribution of PBH to the electromagnetic radiation of the universe, we have to average over the whole Gaussian distribution of relative velocities. The Bondi accretion rate is proportional to $v_{rel}^{-3}$ (see above), and therefore smaller velocities dominate. For this particular case \cite{2017PhRvD..95d3534A} propose to replace the quadratic average of relative velocity and sound speed in Bondi's formula (3.3) above with their harmonic mean: 
\begin{equation}
    v_{eff} = \sqrt{ \langle v_{rel} \rangle~c_s.}
\end{equation}
This is the assumption I adopt here, and the resulting effective velocity $v_{eff}$ is shown as solid red curve in figure \ref{fig:TempVel} (right). With equation (3.8) the accretion rate becomes 
\begin{equation}
    \dot M = 2 \lambda \pi \rho  (G M)^2 ~(\langle v_{rel} \rangle~c_s)^{-3/2}
\end{equation}

\vskip 0.1 truecm
It is interesting to note that in the range 200<z<20 both relative velocity and sound speed decrease linearly with (1+z). Therefore the mass accretion rate is expected to be constant in this era. It is important to understand that the redshift, at which both the sound speed and the relative velocity of the gas turn around due to re-ionization and virialization, respectively, and rapidly increase towards lower redshift, is crucial for our analysis. The redshift, where the minimum velocity occurs, ultimately determines the maximum flux contribution of PBH accretion to the cosmic backgrounds. 

\vskip 0.1 truecm
The calculation of the Bondi accretion rate in equation (3.9) requires the density $\rho$ as a function of redshift. With $\Omega_{bar}$=0.049 and $\rho$=n{$\cdot$}m$_H $, where $n$ is the number density of particles, I find
\begin{equation}
    n = 250~\left( \frac {1+z} {1000} \right)^3~{\rm cm}^{-3}.
\end{equation}
I define $\dot m$ as the normalized mass accretion rate $\dot m = \dot M/\dot M_{Edd} $, with the Eddington accretion rate $\dot M_{Edd}$=1.44$\times10^{17} M/M_\odot$ g s$^{-1}$. Then I can rewrite equation (3.9) into normalized quantities 
\begin{equation}
    \dot m = \lambda \cdot 0.321 \left(\frac {1+z} {1000} \right)^3~\left(\frac {M} {M_\odot} \right) \left( \frac {v_{eff}} {1~{\rm km~s}^{-1}} \right)^{-3}
\end{equation}

\vskip 0.1 truecm
With a very broad PBH mass spectrum, including intermediate-mass and supermassive black holes (M$_{PBH}$>1000 M$_\odot$), it is important to include the effective viscosity due to the Hubble expansion in the early universe \citep{2007ApJ...662...53R}. The Bondi radius determines the amount mass captured by the PBH:
\begin{equation}
    r_B={\frac {G~M} {v_{eff}^2}} \approx 1.34\cdot10^{16} \left(\frac {M} {M_\odot} \right) \left( \frac {v_{eff}} {1~{\rm km~s}^{-1}} \right)^{-2} cm.
\end{equation}
This is shown for two different PBH masses in figure \ref{fig:RadIon} (left). The characteristic time scale for accretion is the Bondi crossing time $t_{cr}=r_B/v_{eff}$, which can be compared to the Hubble time $t_H$ at the corresponding redshift. If $t_{cr}<t_H$ there will be stationary accretion, while for Bondi crossing times larger than the Hubble time the accretion is suppressed. For every redshift we can calculate a critical PBH mass M$_{cr}$, below which the steady-state Bondi assumption can be applied. For redshifts z=1000, 200, 20 this critical mass corresponds to $log (M_{cr}/M_\odot)$=5.3, 4.8, 3.4, respectively. At redshifts below z=20 M$_{cr}$ rapidly increases to values above 10$^6$ M$_\odot$. For PBH masses close to and above M$_{cr}$ the Bondi accretion rate can be scaled by the Hubble viscosity loss given in the dashed curve in figure 3 (left) of \citep{2007ApJ...662...53R}.    

\begin{figure*}[htbp]
    \includegraphics[width=.49\textwidth]{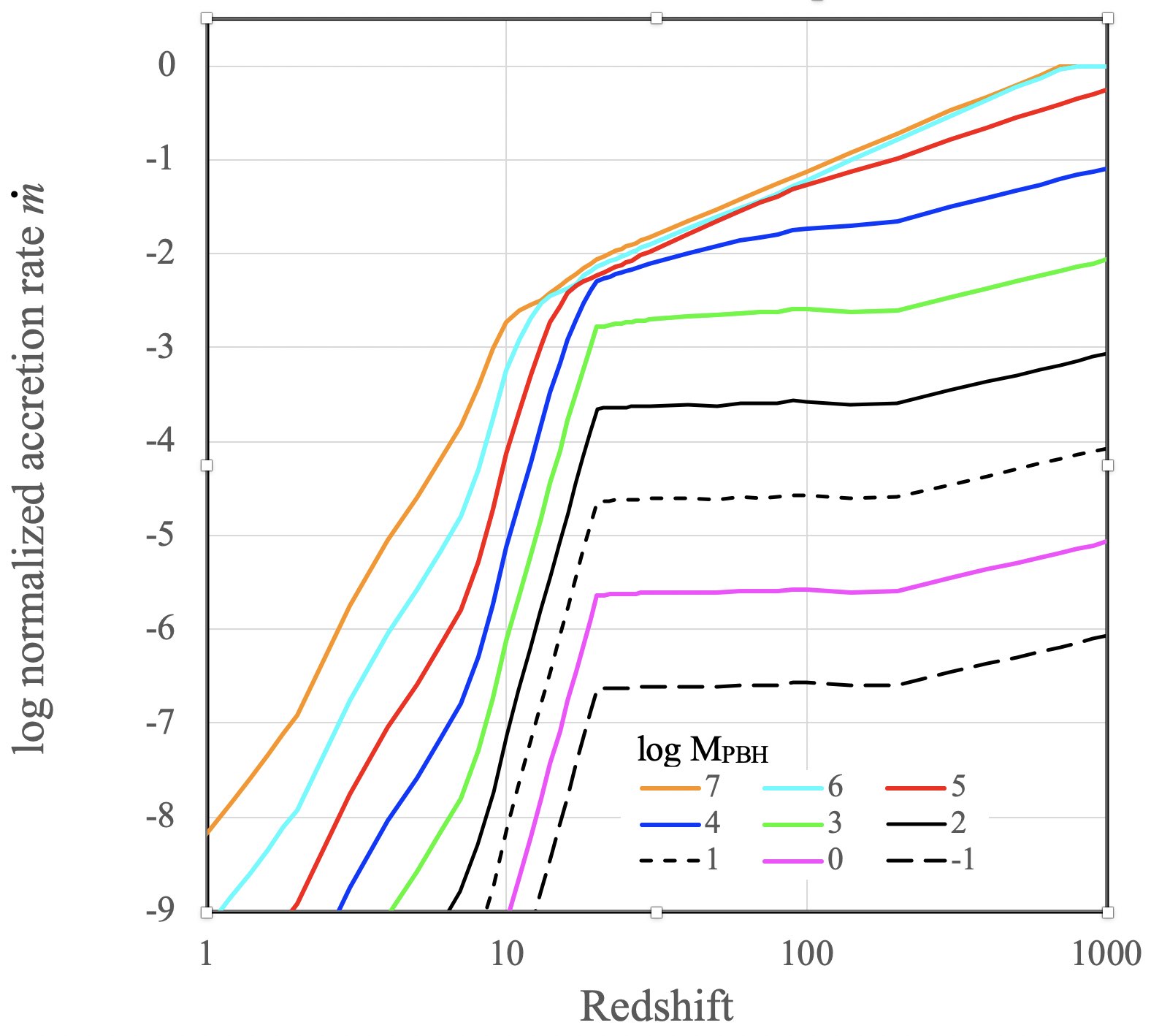}
    \includegraphics[width=.49\textwidth]{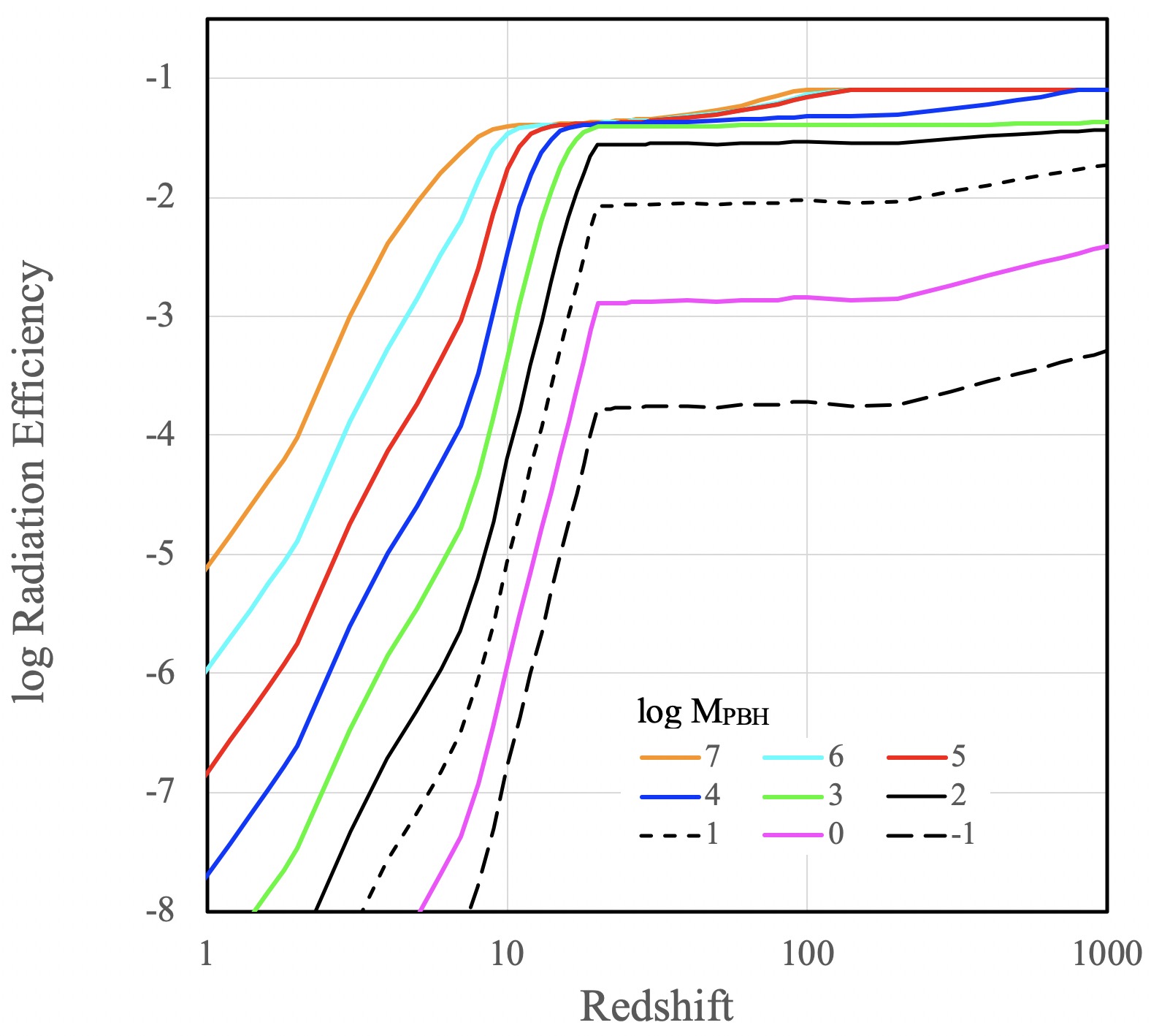}
    \caption{Left: Normalized accretion rate onto PBH with masses in the range 0.1--10$^7$ M$_\odot$ as a function of redshift. Right: Radiative efficiencies derived from the accretion rates, assuming the hot accretion flow model of \cite{2012MNRAS.427.1580X} with a viscous heating parameter $\delta$=0.5.}
    \label{fig:mdoteff}
\end{figure*}

\vskip 0.1 truecm
Inserting $v_{eff}$ from equation (3.8) and figure \ref{fig:TempVel} (right) into equation (3.11), assuming an accretion eigenvalue $\lambda$=0.05 and applying the above Hubble viscosity correction, I can finally calculate the normalized accretion rate as a function of redshift and PBH mass. The results are shown in figure \ref{fig:mdoteff} (left). For PBH masses smaller than $\sim$1000 M$_\odot$ the normalized accretion rate is roughly constant in the redshift range 20<z<200 due to the fact that the density and velocity dependence on redshift in equation (3.9) roughly cancel out (see also the lower panel of figure 4 in \cite{2017PhRvD..95d3534A}). At z<20 $\dot m$ drops dramatically because of the effective velocity increase. PBH masses larger than $\sim10^4$ M$_\odot$ reach accretion rates close to the Eddingon limit at z$\gtrsim$100, but are significantly affected by the Hubble viscosity at z$\gtrsim$20. For all PBH masses the accretion rate is small enough, that the growth of the PBH population can be neglected over cosmic time (PBH with masses in the range 10$^{5-7}$ M$_{\odot}$ accrete about 0.5--2\% of their mass until z>20).

\begin{figure*}[htbp]
    \includegraphics[width=.49\textwidth]{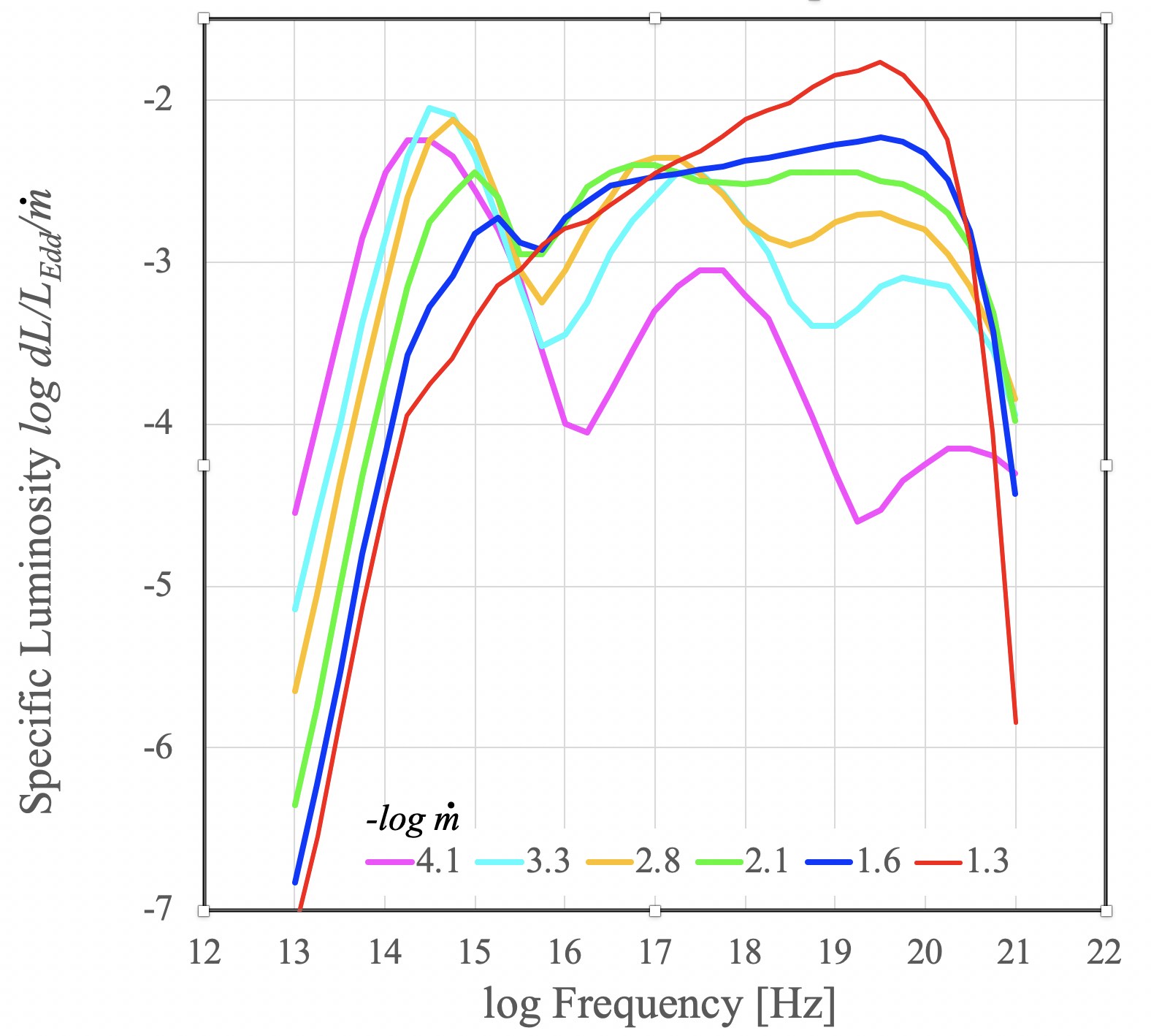}
    \includegraphics[width=.49\textwidth]{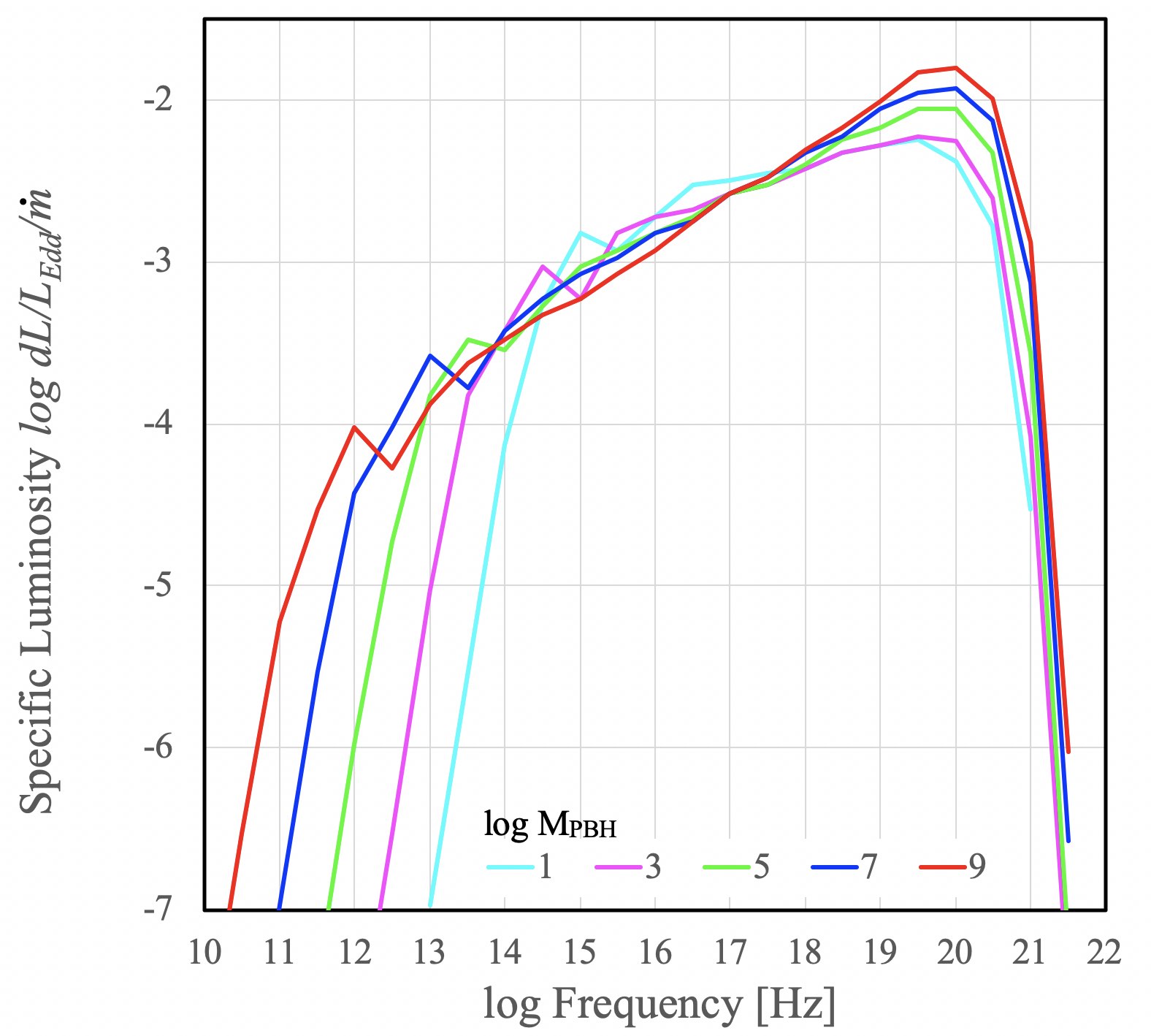}
    \caption{Spectra of the hot disk accretion flow (ADAF) from \citep{2014ARA&A..52..529Y} with a viscous heating parameter $\delta$=0.5, divided by the normalized accretion rate. Left: accretion onto a 10 M$_\odot$ black hole for different accretion rates, as indicated. Right: same for an accretion rate of $log(\dot m)$=-1.6 but different black hole masses (as indicated) .}
    \label{fig:Spectra}
\end{figure*}

\section{Accretion spectrum and radiative efficiency}

For the accretion flow and the electromagnetic emission mechanism I follow \cite{2017PhRvD..96h3524P,2019PhRvD.100d3540M} and assume the formation of an accretion disk. Accretion rates close to the Eddington limit will lead to the formation of a standard Shakura-Sunyaev disk \citep{1973A&A....24..337S}, which has a canonical radiative efficiency $\eta\approx0.1$.  For much lower accretion rates $\dot m$$\ll$1 an advection-dominated hot accretion flow \citep{2014ARA&A..52..529Y} is expected, with a significantly lower radiation efficiency \citep{2012MNRAS.427.1580X}, roughly scaling according to $\eta\propto\dot m$. Figure \ref{fig:Spectra} shows hot accretion flow spectra from \cite{2014ARA&A..52..529Y} with a viscous heating parameter $\delta$=0.5 for black holes, normalized by Eddington luminosity and mass accretion rate. The left graph shows radiation from a 10 M$_\odot$ black hole at different mass accretion rates. The right graph shows the spectrum from black holes with different masses in the range 10-10$^9$ M$_\odot$ and a mass accretion rate $log(\dot m)$=--1.6. 

\vskip 0.1 truecm
It is important to understand, that for advection dominated accretion flows not all the matter entering the Bondi radius will actually reach the black hole. This is due to feed-back mechanisms acting on the accreted gas, e.g. producing outflows or jets. The advection dominated flow models of \cite{2012MNRAS.427.1580X,2014ARA&A..52..529Y} therefore find a radial dependence of mass accretion rate $\dot M\propto R^\alpha$, typically with $\alpha\sim0.4$. Within a radius of about 10 R$_S$, where $R_S=2GM/c^2$ is the Schwarzschild radius, the accretion flow more closely follows the standard Shakura-Sunyaev description of a constant accretion rate with radius down to the last stable orbit ($\sim3R_S$). In terms of the classical Bondi description of quasi-spherical capture, the loss of accreted matter can be associated with the accretion eigenvalue: 
\begin{equation}
    \lambda\approx \left( \frac {10 R_S} {R_D} \right)^\alpha, 
\end{equation} 
where $R_D$ is the outer radius of the accretion disk formed. For $\alpha$=0.4, the value of $\lambda$=0.05 chosen for the analysis in this paper therefore corresponds to an outer disk radius of $R_D\sim$2$\times$10$^4$ $R_S$, about 8 orders of magnitude smaller than the Bondi radius. In this picture the accretion flow on large scales follows the Bondi quasi-spherical flow for most of the radial distance, until the advection-dominated accretion disk is formed.  

\vskip 0.1 truecm
The radiative efficiency for the ADAF spectra in figure \ref{fig:Spectra} is the integral over these curves and has been calculated through numerical simulations by \cite{2012MNRAS.427.1580X}. For this work I use a digitized version of the highest efficiency curve in their figure 1, with a viscous heating parameter $\delta$=0.5\footnote{Please take note that the definition of $\dot m$ between these authors and the analysis presented here differs by a factor of 10.}. A maximum radiative efficiency of $\eta$$\sim$0.08 is achieved for log($\dot m$)>--1.6. We can now calculate the radiative efficiency for every mass and redshift bin from the normalized accretion rate in figure \ref{fig:mdoteff} (left). The result is shown in figure \ref{fig:mdoteff} (right). It turns out that above redshifts z$\gtrsim$20 and black hole masses M$_{PBH}$>100 M$_\odot$, which dominate the contribution to the extragalactic background light, the radiative efficiencies are relatively large (typically >3\%).

\begin{figure*}[htbp]
    \includegraphics[width=.49\textwidth]{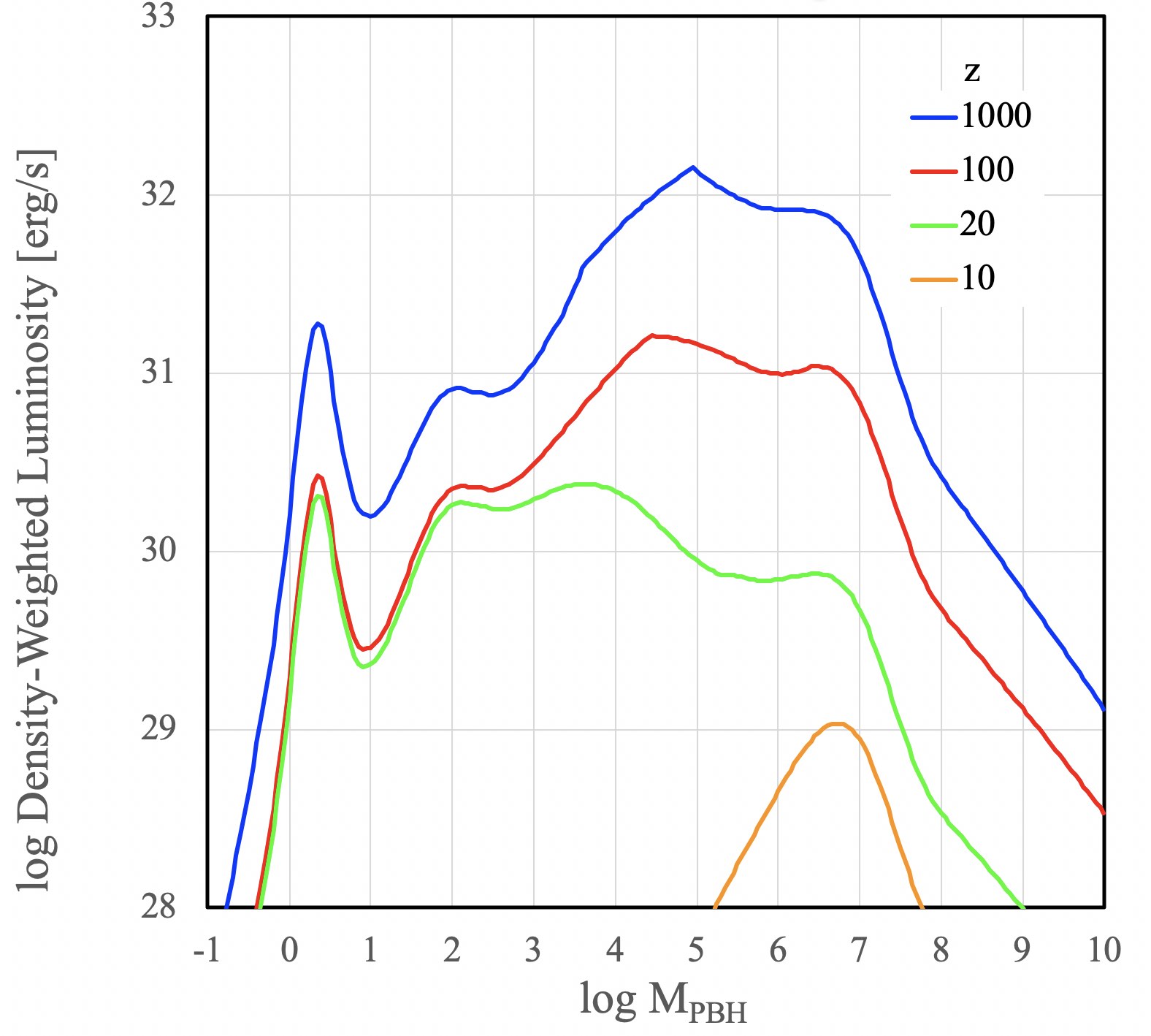}
    \includegraphics[width=.49\textwidth]{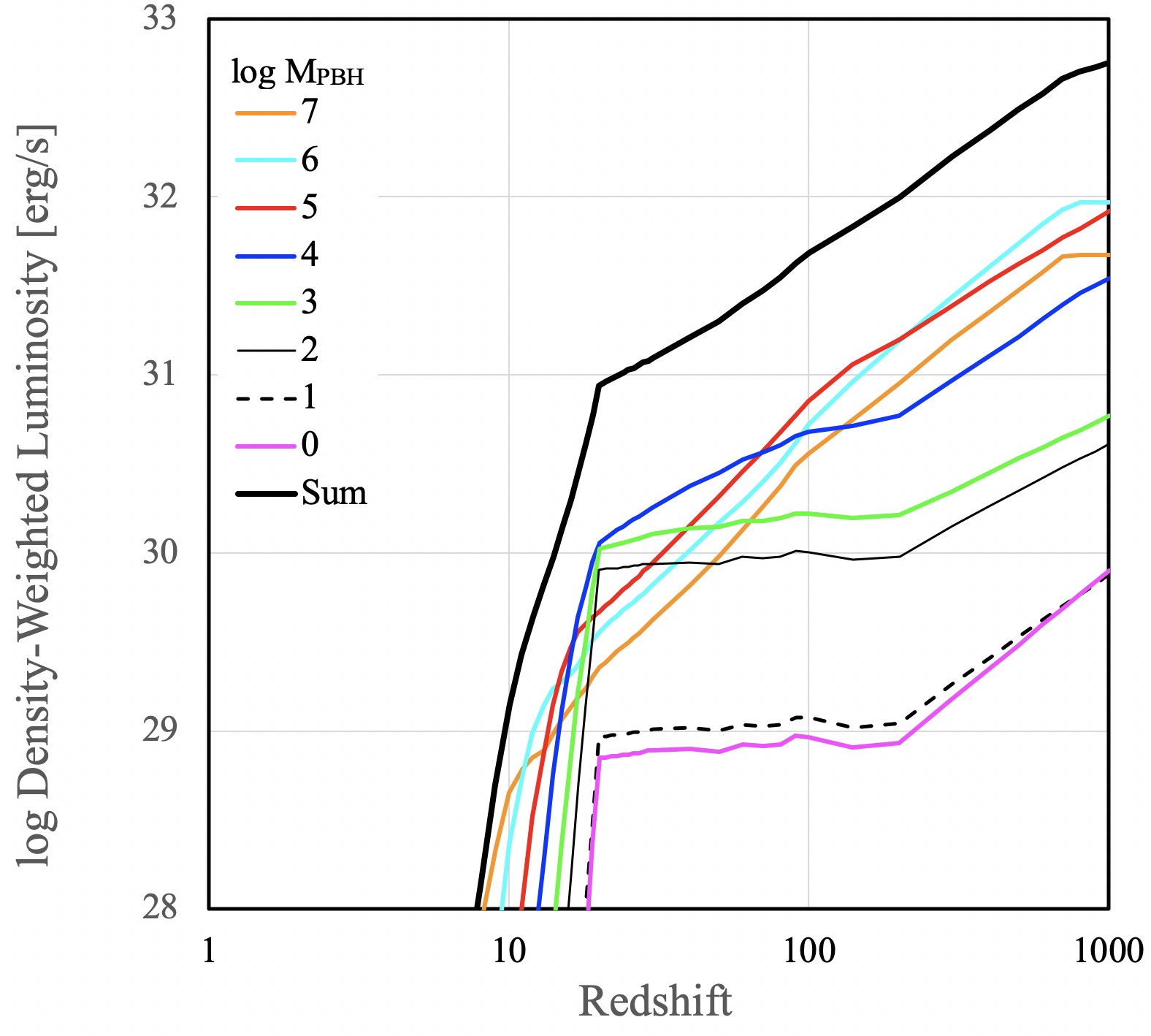}
    \caption{Density-weighted bolometric luminosity of single PBH as a function of mass for different redshifts indicated (left), and as a function of redshift for different mass bins indicated (right).}
    \label{fig:Lbol}
\end{figure*}

\begin{figure*}[htbp]
    \includegraphics[width=.49\textwidth]{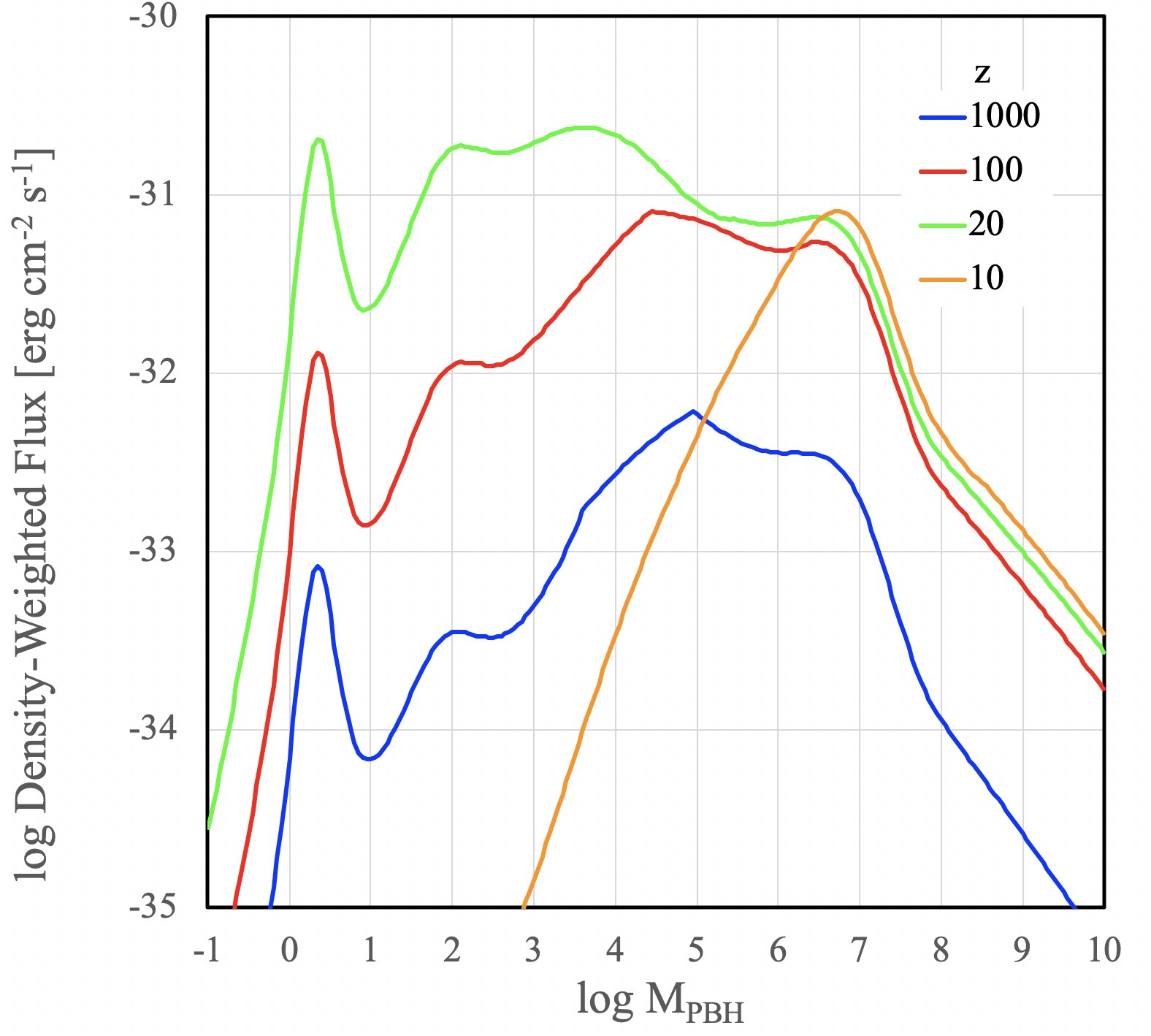}
    \includegraphics[width=.49\textwidth]{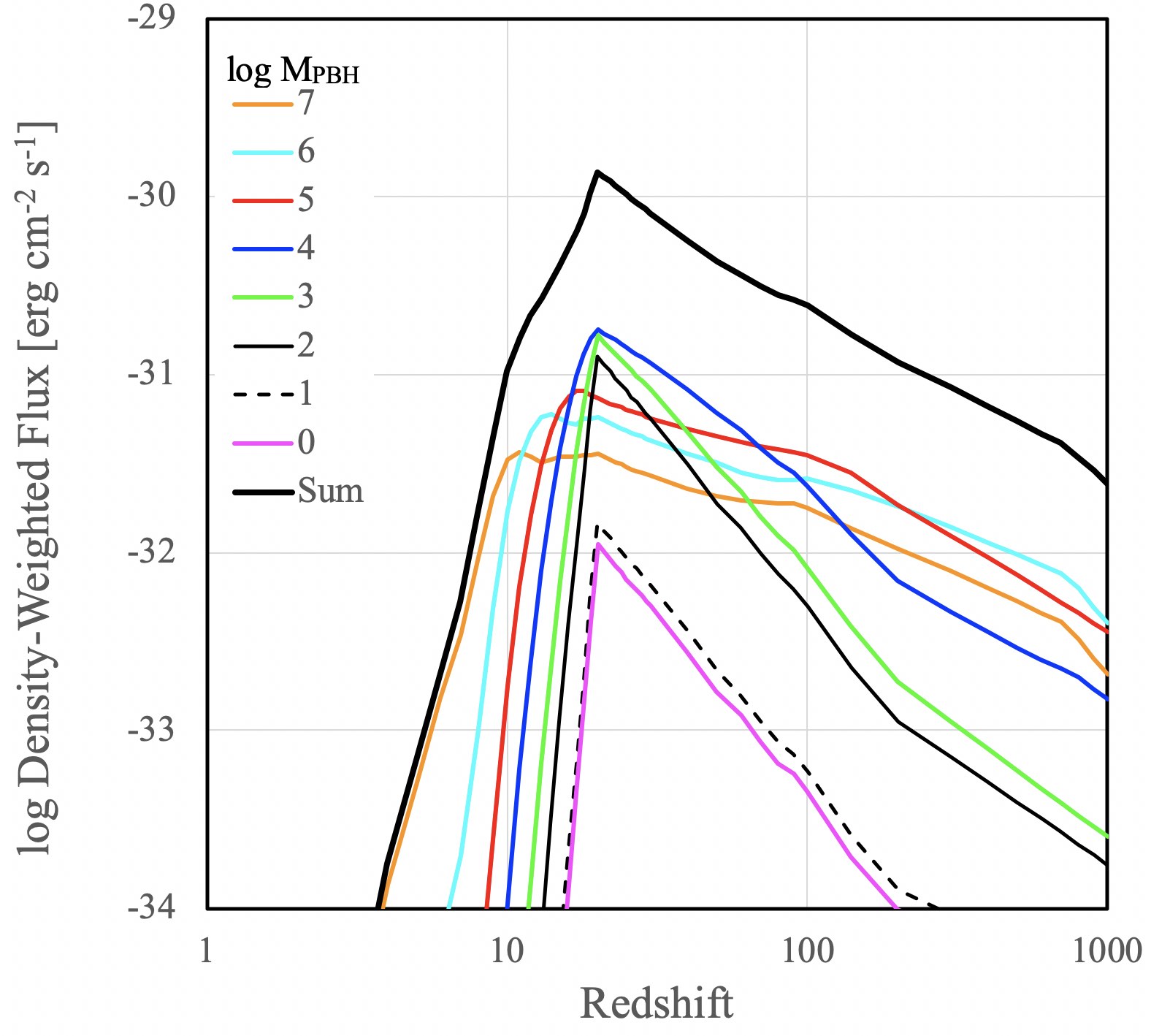}
    \caption{Density-weighted bolometric flux of single PBH as a function of mass for different redshifts indicated (left), and as a function of redshift for different mass bins indicated (right).}
    \label{fig:Fbol}
\end{figure*}

\vskip 0.1 truecm
We now have the ingredients to calculate the bolometric luminosity and flux expected from the baryon accretion onto the assumed PBH mass spectrum over cosmic time. For every black hole of mass M$_{PBH}$ I calculate the expected bolometric luminosity L$_{bol}=\dot m~\eta~L_{Edd}$, where L$_{Edd}$=1.26$\times$10$^{38}~M_{PBH}/M_{\odot}$ erg/s is the Eddington luminosity, and the normalized mass accretion rate $\dot m$ as well as the radiation efficiency $\eta$ are taken from the data in figure \ref{fig:mdoteff}. In every mass bin, the relative number density of PBH compared to those of 1 M$_{\odot}$ is n$_{PBH}$=f$_{PBH}$/M$_{PBH}$, where f$_{PBH}$ is the PBH mass function from figure \ref{fig:fPBH}. For every mass and redshift bin I thus multiply the bolometric luminosity with this relative number density in order to obtain the density-weighted luminosity $\langle L_{bol} \rangle^*$ for an equivalent PBH of 1 M$_\odot$. This quantity is shown in figure \ref{fig:Lbol} as a function of PBH mass (left) and redshift (right). It shows that the largest contribution to the PBH luminosity over cosmic time comes from PBH in the mass range M$_{PBH}$=10$^{3-7}$ at redshifts z>100. The Chandrasekhar PBH mass peak is subdominant in this representation. The total PBH luminosity deposited in the baryonic gas at high redshifts is important for the pre-ionization and X--ray heating of the intergalactic medium discussed in section 6.

\vskip 0.1 truecm
To calculate the contribution of PBH accretion to the extragalactic background light we need to convert the density-weighted luminosities in Figure \ref{fig:Lbol} to bolometric fluxes using the luminosity distance $D_L$ at the corresponding redshift: $\langle F_{bol} \rangle^*$=$\langle L_{bol} \rangle^*$/(4$\pi~D_L^2$). This quantity is shown in figure \ref{fig:Fbol} as a function of PBH mass (left) and redshift (right). It shows that the largest contribution to the extragalactic surface brightness is produced at a redshift z$\approx$20 from PBH in the mass range M$_{PBH}$=10$^{2-5}$, and a similar contribution from the Chandrasekhar mass peak. SMBH at M$_{PBH}\sim10^{6.5}$ have a notable contribution around z$\sim$10.

\section{The contribution of PBH to the extragalactic background light}

\begin{figure*}[htbp]
    \includegraphics[width=.49\textwidth]{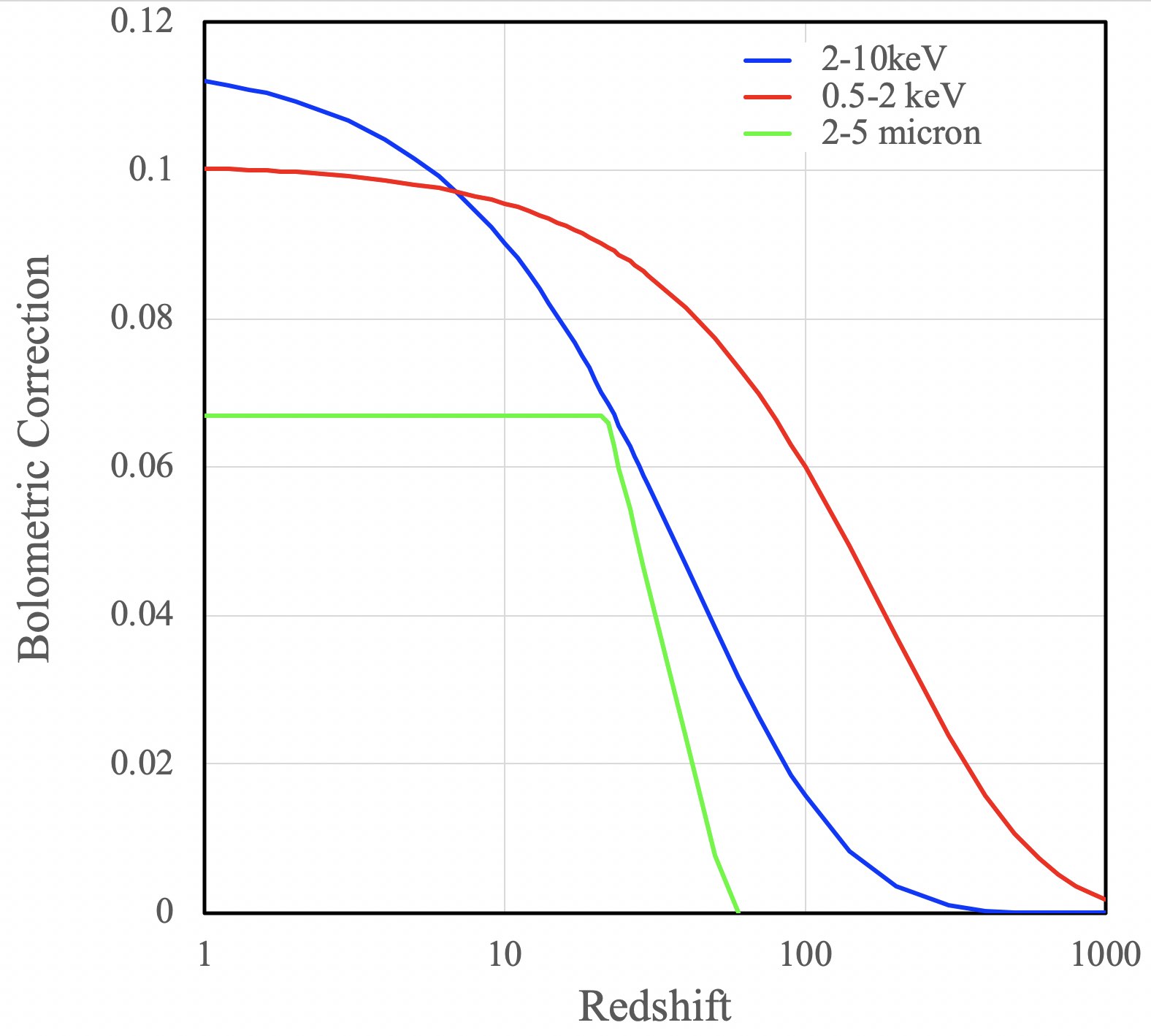}
    \includegraphics[width=.49\textwidth]{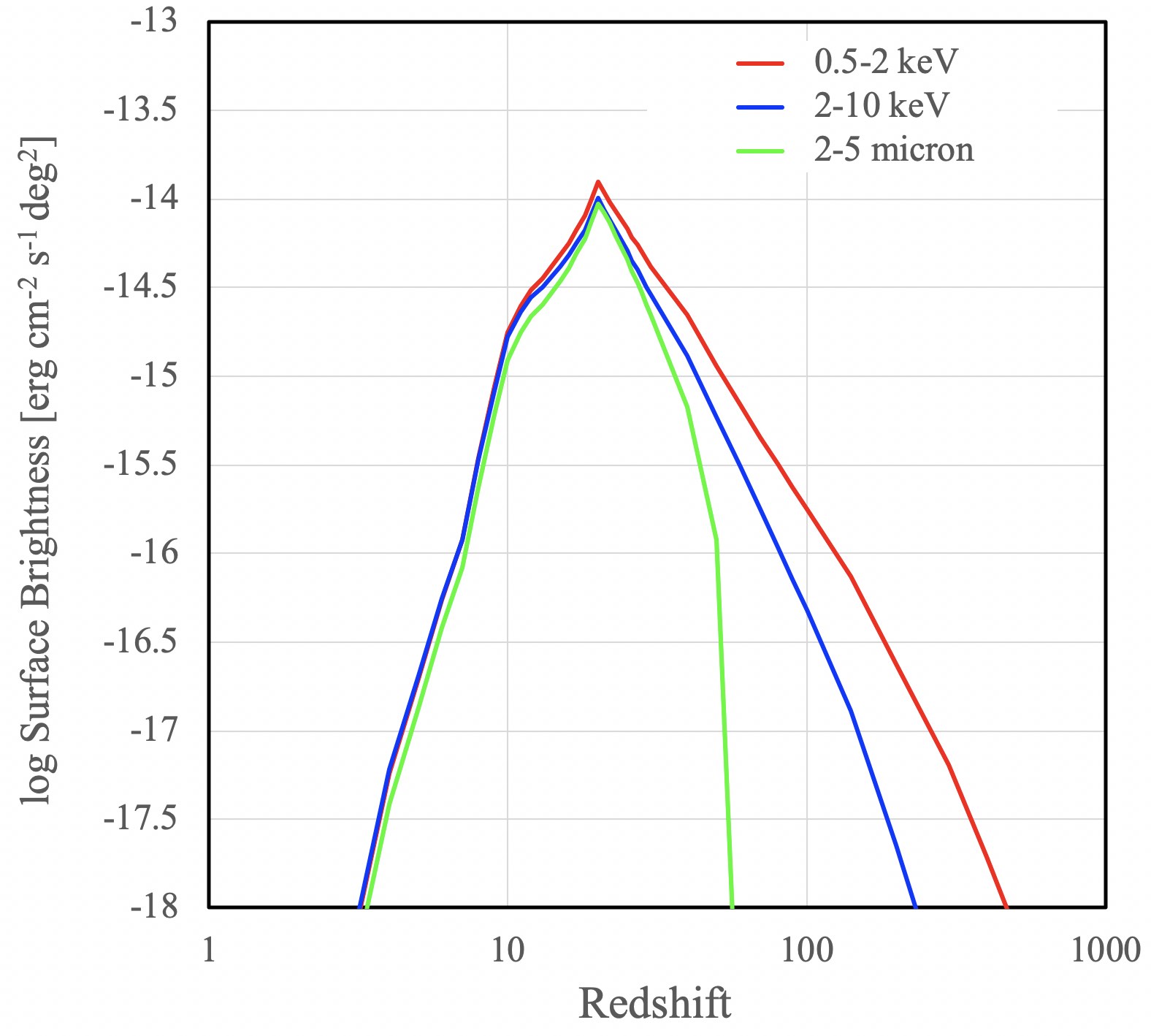}
    \caption{Left: The bolometric correction, i.e. the fraction of the total luminosity falling into the respective observed frequency band as a function of redshift, for the 2--5$\mu$m NIR band, as well as the 0.5--2 and 2--10 keV X--ray bands. Right: Predicted surface brightness of the PBH in the same observed bands as a function of redshift (per $\Delta$z=1). }
    \label{fig:kcorrSB}
\end{figure*}

\vskip 0.1 truecm
To calculate the surface brightness per redshift shell in a particular observed frequency band [$\nu_1$;$\nu_2$] of the electromagnetic spectrum, I take into account the spectral shape and the fraction of the radiation falling into the rest frame frequency band [$\nu_1$/(1+z);$\nu_2$/(1+z)]. The exact spectral shape is not so important for this derivation, it is mainly used to calculate the bolometric corrections, i.e. the fraction of the total luminosity falling into the various frequency bands as a function of redshift. The ADAF spectra in figure \ref{fig:Spectra}, in particular those at high $\dot m$ values, can be approximated by power laws with an exponential cutoff at $\sim$200 keV. Following \cite{2017PhRvD..96h3524P} and \cite{2019PhRvD.100d3540M}, I assume a power law slope of --1 (corresponding to a flat line in figure \ref{fig:Spectra}). Below a critical frequency $\nu_c$ the power law spectrum is cut off by synchrotron self-absorption into a power law with a steep slope of approximately +1.86. As can be seen in figure \ref{fig:Spectra} (right), $\nu_c$ depends on M$_{PBH}$ and can be approximated by log($\nu_c$)$\approx$14.82--0.4log(M$_{PBH}$/M$_\odot$). The bolometric corrections are then obtained by integrating the analytic normalized spectra over the observed frequency bands. For the 2--5$\mu$m band we have to consider in addition the Lyman-$\alpha$ break, which produces a sharp cutoff at z$\gtrsim$30 (see e.g. \cite{2013MNRAS.433.1556Y,2002MNRAS.336.1082S}). These bolometric corrections are shown in figure \ref{fig:kcorrSB} (left) for the 2--5$\mu$m NIR band, the 0.5--2 keV and the 2--10 keV X--ray bands, respectively. 

To predict the surface brightness of all PBH across cosmic time in these observed frequency bands, the total flux per PBH in figure \ref{fig:Fbol} (right) has to be multiplied with the bolometric correction and the PBH surface density in a particular redshift shell. Using the critical mass density of the universe $\rho_{crit}$=1.26$\times10^{20} M_\odot~{\rm Gpc}^{-3}$ and the Dark Matter density $\Omega_{DM}$=0.264, as well as the reference mass 1M$_\odot$, a comoving PBH space density of $n_{PBH} = 3.32\times10^{19} (1+z)^3~{\rm Gpc}^{-3}$ is obtained. For every redshift shell [z+$\Delta$z] the PBH space density is multiplied with the volume of the shell [V(z+$\Delta$z)--V(z)] and divided by 4$\pi$ deg$^2$ to obtain the number of PBH per deg$^2$. Figure \ref{fig:kcorrSB} (right) shows the derived surface brightness as a function of redshift (per $\Delta$z=1 interval) for the three spectral bands considered here. The emission in all three bands peaks around z$\approx$20 with a FWHM of $\Delta$z$\approx$[-3;+6].

\vskip 0.1 truecm
The curves in figure \ref{fig:kcorrSB} (right) can now be integrated to predict the total PBH contribution to
the extragalactic background light as SB$_{2-5\mu}$$\approx$10$^{-13}$,
SB$_{0.5-2keV}$$\approx$1.9$\times$10$^{-13}$, and SB$_{2-10keV}$$\approx$1.3$\times$10$^{-13}$ erg cm$^{-2}$
s$^{-1}$ deg$^{-2}$, respectively. 
The minimum amount of X--ray surface brightness necessary to explain the CXB/CIB cross-correlation signal observed by \cite{2013ApJ...769...68C} in the 0.5--2 keV band has been discussed by \cite{2019MNRAS.489.1006R}. This is $9 \times 10^{-14}$ erg cm$^{-2}$ s$^{-1}$ deg$^{-2}$, corresponding to roughly 1\% of the total CXB signal in this band. The 0.5--2 keV PBH contribution predicted for an accretion eigenvalue of $\lambda$=0.05 in equation (3.11) is thus about a factor of 2 larger than the observed CXB fluctuation signal, which could well be consistent, given the coherence between the CXB and CIB signals. As discussed above, there is a marginally significant diffuse CXB remaining after accounting for all discrete source contributions \citep{2006ApJ...645...95H,2017ApJ...837...19C}. Extrapolating into the X--ray bands considered here, this residual flux corresponds to $\approx$(7$\pm$3) and $\approx$(9$\pm$20)$\times10^{-13}$ erg cm$^{-2}$ s$^{-1}$ deg$^{-2}$ in the 0.5--2 keV and 2--10 keV band, respectively. Assuming the $\lambda$=0.05 value, the predicted PBH contribution is therefore well below the upper limit (15--25\%) of any unresolved component left in the CXB. {\bf The main result of this paper is therefore, that the assumed PBH population for the dark matter can indeed explain the X--ray fluctuation signal, with a Bondi accretion eigenvalue of $\lambda$=0.05.}

\vskip 0.1 truecm
The flux measured in the 2--5$\mu$m CIB fluctuations at angular scales >100" is about 1 nW m$^{-2}$ sr$^{-1}$ \citep{2007ApJ...654L...1K}, or 3$\times10^{-10}$ erg cm$^{-2}$ s$^{-1}$. The cross-correlated CIB/CXB fluctuations contribute about 10\% to the total CIB fluctuations \citep{2013ApJ...769...68C}, i.e. 3$\times10^{-11}$ erg cm$^{-2}$ s$^{-1}$. Therefore the predicted PBH contribution to these CIB fluctuations is only about 0.5\% for $\lambda$=0.05. It is argued in \cite{2016ApJ...823L..25K} that PBH in the early universe could amplify the cosmic power spectrum at small spatial scales (see below). Together with the pre-ionization of the intergalactic medium discussed below, the PBH can therefore significantly increase the associated star formation. The NIR emission in this picture would then be dominated by early star formation associated with PBH instead of direct PBH emission.

\section{Discussion}

\subsection{Linear versus post-linear growth}

In this simplified treatment I only consider the linear evolution of the power spectrum above the virialization redshift around z$\approx$20 (see figure \ref{fig:TempVel} right). On sufficiently large scales the initial power spectrum has been very precisely determined as nearly scale invariant with overdensities of 10$^{-4}$ \cite{2018arXiv180706209P}, and the PBH density field is expected to follow the standard adiabatic perturbations. On small scales the power spectrum is only poorly constrained and could be significantly amplified by the discrete nature of the PBH population itself \cite{2016ApJ...823L..25K,2018MNRAS.478.3756C,2019PhRvD.100h3528I}. Poisson variations in the density of PBH will introduce non-linear growth of density fluctuations and the corresponding velocity dispersion already well before the virialization redshift z$\sim$20 discussed above. However, from numerical simulations \cite{2019PhRvD.100h3528I} conclude that the nonlinear velocity perturbations introduced by >20 M$_\odot$ PBH are too small to dominate the relative velocities between baryons and PBH at z$\gtrsim$100 \citep[see also][]{2019PhRvD.100h3016H}. However, non-linear effects definitely become more important at lower redshifts (see above) and could effectively reduce the Bondi capture rate.

\subsection{Magnetic fields in the early universe}

The accretion mechanism assumed in the Bondi capture model only works, if there is a rough equipartition between the kinetic energy and magnetic fields in the accreted gas, as it is the case in the turbulent interstellar medium of our Galaxy. It is therefore justified to ask, whether this mechanism can also work at high redshifts, where the existence and magnitude of magnetic fields is still unclear. Magnetic fields are present at almost every scale of the low redshift universe, from stars and planets to galaxies and clusters of galaxies and possibly even in the intergalactic medium in voids of the cosmic web, as well as in high-redshift galaxies. \cite{2001PhR...348..163G} and \cite{2016RPPh...79g6901S} review the observations and possible origin of magnetic fields. There is a surprising similarity between the relatively strong magnetic fields measured in our own Galaxy (0.3--0.4 nT) and other nearby galaxies ($\sim$1 nT) with magnetic fields measured in clusters of galaxies (0.1--1 nT), as well as in high redshift galaxies ($\sim$1 nT), when the universe was only about 1/3 of its current age. There are even indications of magnetic fields of order $\gtrsim$10$^{-20}$ T in cosmic voids derived from the gamma ray emission of blazars \citep{2013ApJ...771L..42T}. 

\vskip 0.1 truecm
One can come to the conclusion that the  origin  of  cosmic  magnetism  on  the  largest  scales  of  galaxies, galaxy clusters and the general intergalactic medium is still an open problem \citep{2018arXiv180903543S}. It is usually assumed that primordial or cosmic seed fields are amplified over time through the galactic dynamo effect to produce the rather strong fields observed in local galaxies. In this picture it is, however, unclear how similar fields can be created in such different settings (e.g. clusters) and different cosmic times (high-redshift galaxies). An interesting possibility is therefore that cosmic magnetic fields could be remnants from the early universe, or created in a process without galactic dynamos. 
Assuming equipartition, the energy density in the CMB photons would correspond to a magnetic field of about 0.3 nT. Magnetic fields of $10^{-20}$ T, as observed in cosmic voids today, would only require a minute fraction of $10^{-10}$ of this energy density in the early universe to be channeled into magnetic fields. 

\begin{figure*}[htbp]
    \includegraphics[width=.49\textwidth]{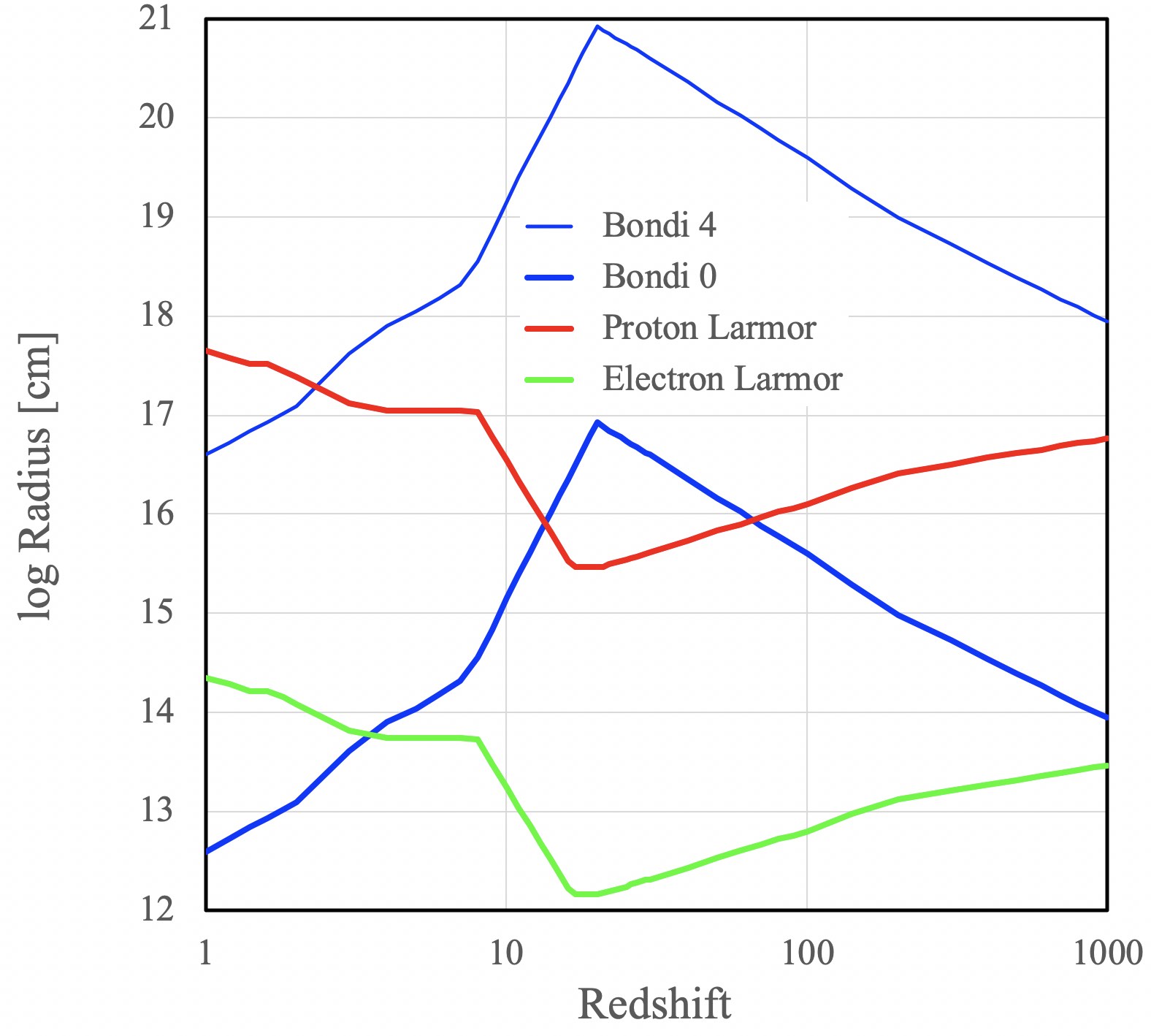}
    \includegraphics[width=.49\textwidth]{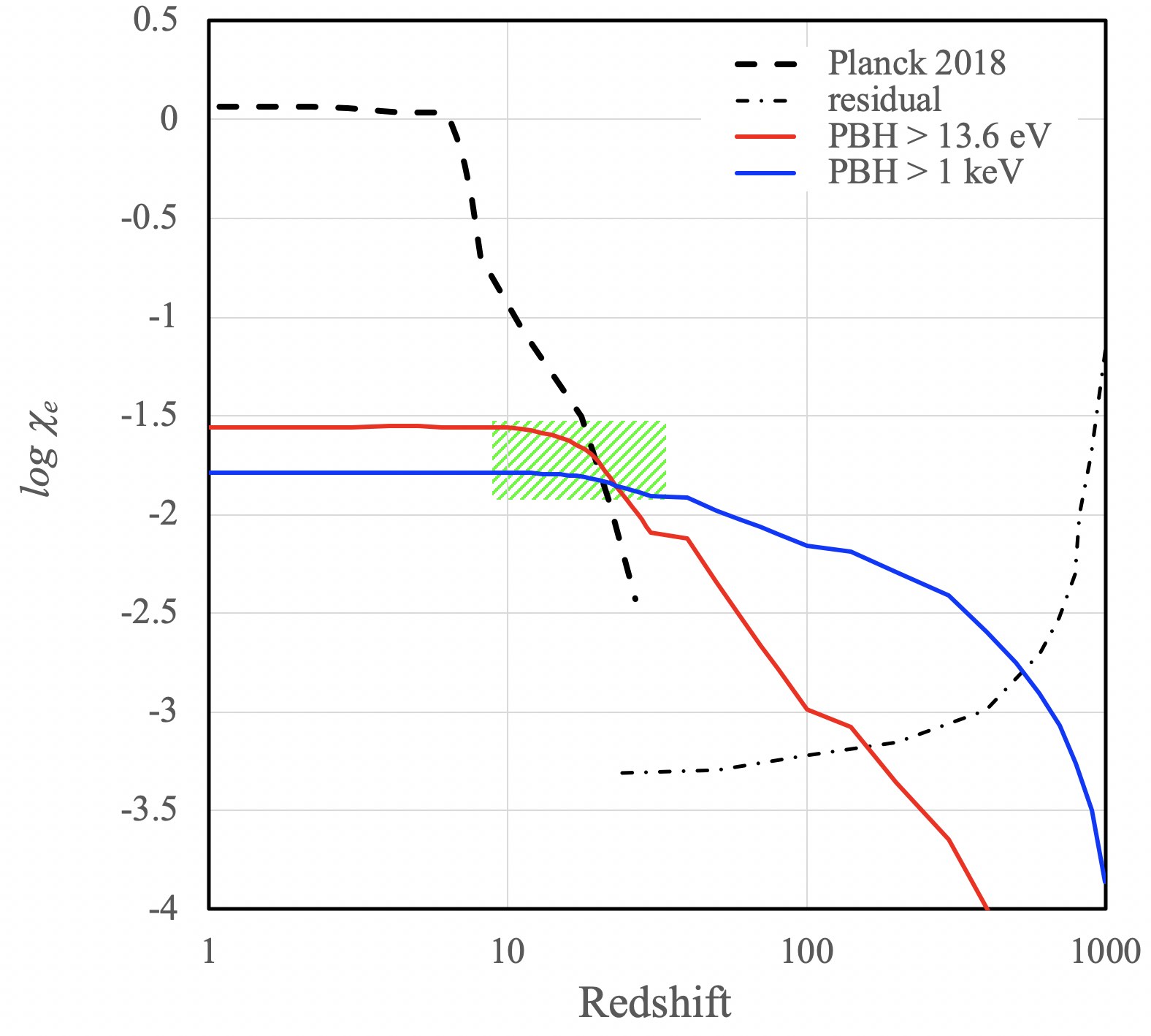}
    \caption{Left: The Bondi radius for a 10$^4$ M$_\odot$ (thin blue) and 1 M$_\odot$ (thick blue) PBH compared to the proton (red) and electron (green) Larmor radius, assuming a magnetic field of B=$10^{-20}$ T, as observed in local galaxy voids. Right: Baryon ionization/heating fraction $\chi_e$ as a function of redshift. The thin dash-dotted line shows the residual ionization left over from the radiation dominated era \cite{2002MNRAS.330L..43B}. The red curve shows the ionization fraction from UV photons produced by accreting PBH. The blue curve shows the corresponding heating fraction by >1 keV X--ray photons. The thick dashed black line shows one of the models consistent with the {\em Planck} satellite data \citep{2018arXiv180706209P} (see text). The green hatched areas shows the range of high--redshift ionization fractions considered in \cite{2018RvMP...90b5006K}.}
    \label{fig:RadIon}
\end{figure*}

\vskip 0.1 truecm
Here I argue, that PBH could play a role in amplifying early magnetic seed fields and sustaining them until the epoch of galaxy formation. I compare the Bondi radius in eq. (3.12) and figure \ref{fig:RadIon} (left) with the Larmor radius
\begin{equation}
    r_L={\frac {m~v_\bot} {|q|~B}},
\end{equation}
which determines the gyro motion of particles moving through a magnetic field. Here $m$ is the mass of the particle (either proton or electron), $v_\bot$ is the velocity component of the particle perpendicular to the magnetic field, $|q|$ is the absolute electric charge of the particle, and $B$ is the magnetic field. Assuming a seed field of $B$=10$^{-20}$ T and approximating the velocity with the sound speed $v_\bot$$\approx$$c_s$ yields the gyro radius for both protons and electrons. The proton gyro radius is about a factor of 2000 larger than the electron gyro radius. 

\vskip 0.1 truecm
Figure \ref{fig:RadIon} (left) shows the Bondi radius as well as the proton and electron Larmor radii as a function of redshift. If the gyro radius is smaller than the Bondi radius, the respective particle is easily accreted by the PBH. If, however, the gyro radius is larger than the Bondi radius, the particle will first not be easily accreted, but rather spiral around the PBH. From  \ref{fig:RadIon} we see, that at redshifts z$\gtrsim70$ and PBH masses in the range M$_{PBH}$$\approx$0.3--500 for our assumed magnetic field strength the proton Larmor radius is larger than the Bondi radius, while the electron Larmor radius is smaller than the Bondi radius. There is still a substantial fraction of residual electrons and protons/helium ions from the era before recombination (see the dash-dotted curve in \ref{fig:RadIon} right from \cite{2002MNRAS.330L..43B}). These electrons have therefore no problem being accreted, while for certain PBH masses protons resist the accretion. This will create a net electric current, which in turn will increase the average magnetic field strength until the proton gyro radius becomes smaller than the Bondi radius. This way the PBH can amplify the average magnetic field. The  supersonic  motion between baryon gas and PBH discussed above is expected to be coherent over large scales (of the order of Mpc) and can therefore induce large-scale ordered magnetic fields. A further magnetic field amplification occurs, as discussed above, in the accretion funnel, when magnetic fields are dissipated through reconnection and ejected with the plasma. In a sense, the ubiquitous PBH can possibly create their own magnetic field and distribute it throughout the universe. It is, however, plausible to assume, that magnetic fields in the early universe should be smaller than today, and the fractions of ionized baryons is less. This could also explain a rather small Bondi accretion eigenvalue required to match the observations.      

\subsection{Re-Ionization}
\label{subsec:Reionization}

Next I turn to the contribution of PBH accretion to the re-ionization and re-heating history of the universe. At z$\approx$1089, when the photons decoupled from the baryons and formed the CMB radiation, the universe became predominantly neutral. Afterwards the universe entered a long period of “darkness”, in which the residual ionization left over from the primordial photon-baryon fluid diminished (see figure \ref{fig:RadIon} right), the background photons cooled down, and any higher-frequency emission was quickly absorbed in the atomic gas. In the model described here the first sources to illuminate the “dark ages” would be the PBH accreting from the surrounding gas. Their ultraviolet emission, above the Hydrogen ionization energy of 13.6 eV, would start to re-ionize small regions around each PBH. However, in the beginning the density of the surrounding gas is still so high that the ionized matter quickly recombines. As long as the re-combination time is much shorter than the Hubble time at the corresponding epoch, UV photons from the PBH cannot penetrate the surrounding medium, but instead produce an ionized bubble growing with time. In this type of ionization equilibrium the number of ionizing photons $N_{ion}$ required to overcome recombination is given as the ratio between the Hubble time $t_H(z)$ and the recombination time $t_{rec}(z)$ at the particular epoch, and can be derived from equations (2) and (3) in \cite{2004MNRAS.350..539R} as
\begin{equation}
    N_{ion} = t_H/t_{rec} = max[1, 0.59~\left( \frac {1+z} {7} \right)^{1.5}].
\end{equation}
At a redshift z=1000, $N_{ion}$ is about 1000, and reaches a level of unity at z$\lesssim$10 for the assumed set of parameters. For this calculation I ignore clumping of the ionized gas. In reality the effective clumping factor is relatively large for reionization at high redshift because the ionizing sources are more numerous in the filaments of the cosmic web, but must also ionize a correspondingly larger fraction of dense gas in the filaments, and thus ionization is slowed down. At lower redshift, when molecular gas and stars have already formed, not all UV photons will escape the dense regions. The effective escape fraction is one of the largest uncertainties in our current understanding of re-ionization. For simplicity, I assume an escape fraction $f_{esc}$=0.1 for UV photons, and $f_{esc}$=1 for X--ray photons, independent of redshift.

\vskip 0.1 truecm
To calculate the history of pre-ionization by PBH I integrate the above normalized ADAF model for frequencies log($\nu$)>15.52 Hz, corresponding to the hydrogen ionization energy of 13.6 eV. To calculate the number of ionizing photons per PBH of reference mass 1 M$_\odot$ I take the density-weighted luminosity $\langle L_{bol} \rangle^*$ from figure \ref{fig:Lbol} (right). To determine the average space density of ionizing photons I multiply with the average comoving space density of PBH (assuming the reference mass 1 M$_\odot$):
\begin{equation}
    n_{PBH} =  1.06\times10^{-54}~\left( \frac {1+z} {1000}  \right)^3~{\rm cm}^{-3}, 
\end{equation}
and with the escape fraction $f_{esc}$ and finally divide by $N_{ion}$ from eq. (6.3) and the average density of baryons given in equation (2.10) to determine the ionization rate of baryons over cosmic time. 

\vskip 0.1 truecm
The red curve in figure \ref{fig:RadIon} (right) shows the cumulative ionization fraction $\chi_e$ as a function of redshift for the accretion eigenvalue $\lambda$=0.05. A maximum cumulative ionization fraction of $\sim$2.8\%, is reached at a redshift z$\approx$10. This can be compared to one of the recent models determined from the {\em Planck} satellite data \citep{2018arXiv180706209P}. Here the 1$\sigma$ upper percentile of the {\em FlexKnot} model in their figure 45, which is consistent with the ionization optical depth determined from the most recent {\em Planck} data, is shown as dashed curve. A high-redshift contribution to the ionization history of the universe has also been discussed by \cite{2018PhRvD..98f3514H} and \cite{2018RvMP...90b5006K}. The range of $\chi_e$ values assumed in the latter work is shown as green hatched region in figure \ref{fig:RadIon} (right). For the choice of $\lambda$=0.05, the UV emission from the PBH population assumed in the toy model is therefore fully consistent with the observational constraints from {\em Planck}.

\subsection{X--ray heating}

The role of X--ray heating in shaping the early universe has been discussed by \cite{2013MNRAS.431..621M}. Compared to UV photons, X--ray photons have a much larger mean free path and can therefore ionize regions far away from the source. In addition, most of the X--ray energy gets deposited into heating up the gas. In order to estimate the amount of X--ray heating of the gas I applied the same mechanism as for the UV photons above, but integrating the above ADAF model for frequencies log($\nu$)>17.68 Hz, corresponding to 2 keV. I assume an escape fraction of $f_{esc}$=1 and $N_{ion}$=1. The blue curve in figure \ref{fig:RadIon} (right) shows the cumulative 2 keV heating fraction per baryon as a function of redshift for the assumed accretion eigenvalue of $\lambda$=0.05. The maximum cumulative heating fraction is $\sim$1.6\%. X--rays from PBH therefore have only a small contribution to the pre-ionization of the universe as a whole, but can be responsible for a part of the pre-heating of gas observed in the "entropy floor" of groups of galaxies. \cite{2002A&A...383..450D} reviewed the energetics of groups and clusters of galaxies, which cannot be reproduced by simple models, where the gas  density is proportional to dark matter density. \cite{1991ApJ...383..104K} and \cite{1991ApJ...383...95E} argued that the gas must have been pre-heated before falling into the cluster potential. X--ray observations of groups of galaxies with {\em ROSAT} by \cite{1999Natur.397..135P} confirmed the need for a non-gravitational entropy injection in the group gas. These authors coined the term "entropy floor", which amounts to an energy of about 2 keV per baryon injected into the group gas. The pre-heating of the gas by PBH, albeit only contributing to a small fraction of the total baryon content of the universe, could have played an important role in heating the high-density regions, which first formed groups and clusters. 

\subsection{Cosmological 21-cm signal}

\begin{figure*}[htbp]
\begin{center}
    \includegraphics[width=.49\textwidth]{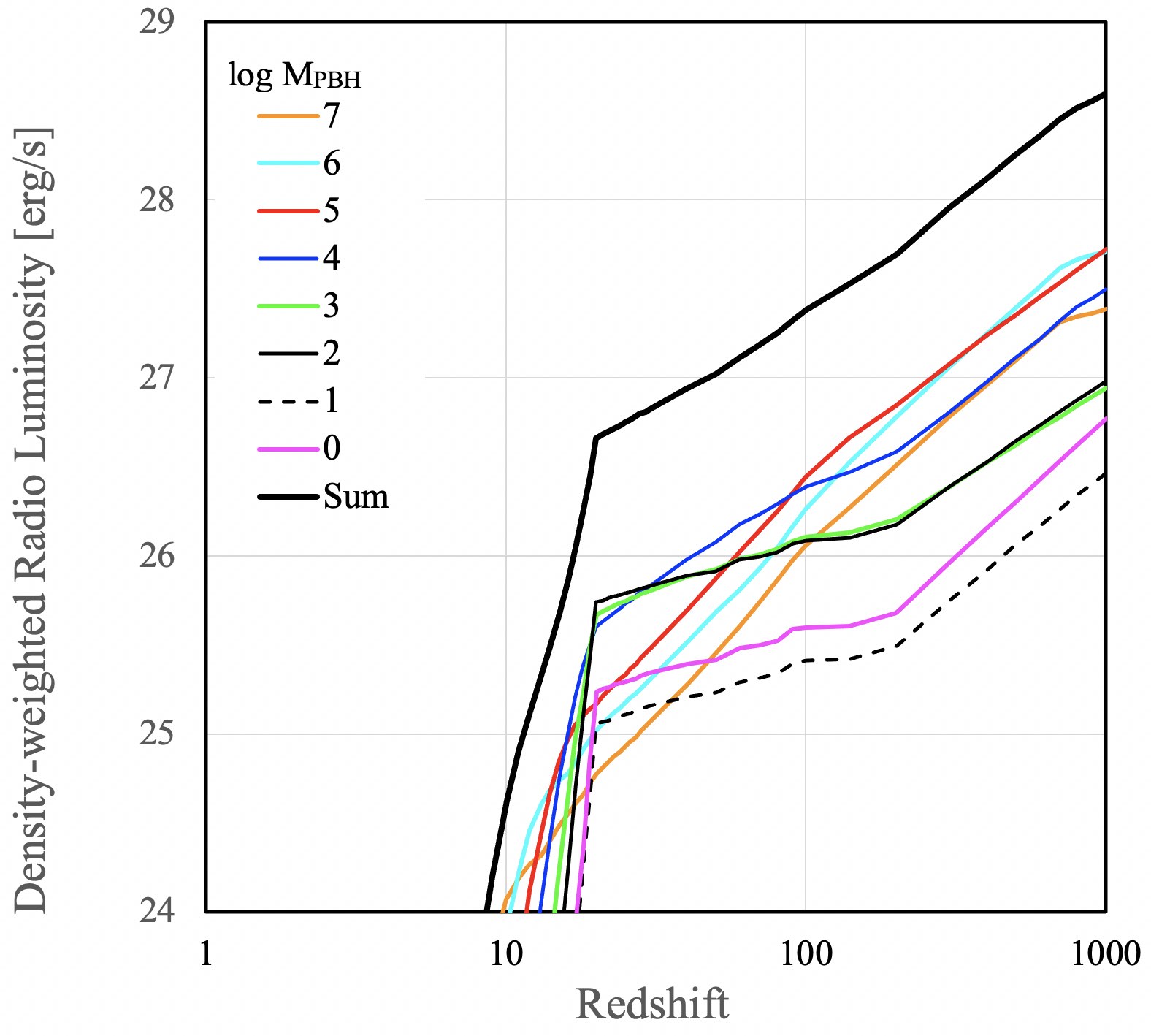}
    \caption{Density-weighted 1.4 GHz (observed) luminosity of a single PBH as a function of mass for different redshifts indicated.}
    \label{fig:LRadio}
\end{center}
\end{figure*}

The red-shifted 21-cm line can provide important new constraints on the physical processes in the early universe \citep[see e.g.][]{2016ApJ...821...59F,2019PhRvD.100d3540M}. The Experiment to Detect the Global EoR Signature ({\em EDGES}) has measured a strong, sky-averaged 21-cm absorption line profile after subtracting the Galactic synchrotron emission \citep{2018Natur.555...67B}. The signal is centered  at a frequency around 78 MHz and covers a broad range in redshift z=15--20. The signal may be due to ultraviolet light from first objects in the universe altering the emission of the 21-cm line by lowering the spin temperature of neutral hydrogen relative to the CMB. However, the signal is about three times larger than that expected from the standard $\Lambda$CDM cosmology, which led some authors to suggest new dark matter physics \citep[e.g.][]{2018PhRvL.121a1101F}. Instead of new physics, an increased 21-cm radio background contribution above the CMB at the epoch around z=15--20 could also explain the {\em EDGES} signal. Indeed, \cite{2018ApJ...868...63E} estimate the additional 21-cm radio background from accretion onto putative radio-loud intermediate-mass black holes (IMBH) forming in first molecular cooling halos at redshifts z=16--30. This could be sufficient to explain the {\em EDGES} feature, however, it requires extreme assumptions about the radio loudness of the IMBH population. Instead of assuming an interpretation in terms of mini-QSOs from IMBH grown through accretion, I estimate here, whether PBH accretion could have a significant contribution to the EDGES signal. 
A full treatment of this effect for the PBH toy model is beyond the scope of this paper, but similar to the treatment of the PBH contribution to the CXB and CIB derived in section 5, I can estimate the PBH contribution to the observed low-frequency cosmic radio background, and thus to the EDGES signal. 

\vskip 0.1 truecm
The balloon-borne double-nulled Absolute Radiometer for Cosmology, Astrophysics and Diffuse Emission ({\em ARCADE2}) instrument has measured the absolute temperature of the sky at frequencies 3, 8, 10, 30, and 90 GHz, and detected a significant excess over the CMB blackbody spectrum at a temperature of 2.731K \citep{2011ApJ...734....5F}. Combining the {\em ARCADE2} measurements with lower frequency data from the literature, the excess brightness temperature can be characterized as a power law T$_R$=1.19 ($\nu$/1 GHz)$^{-2.62}$ K, which translates into a radio spectrum with a slope of -0.62 and a normalization of  3$\times$10$^{-22}$ W m$^{-2}$ sr$^{-1}$ at 1.4 GHz. This cosmic radio synchrotron background is substantially larger than that expected from an extrapolation of the observed radio counts \citep{2018PASP..130c6001S}, and thus presents a major new challenge in astrophysics. \cite{2018ApJ...858L..17F} found that the global 21cm signal can be significantly amplified by an excess background radiation compared to the standard $\Lambda$CDM models, especially in absorption. Assuming that only 10\% of the radio synchrotron background originates at large redshifts, they predict a 21cm feature almost an order of magnitude stronger than that expected purely from the CMB. Interpolating between their calculations for 0.1\% and 10\% excess background I find, that an excess high-redshift radiation field of about 5\% of the observed radio synchrotron background is sufficient to explain the EDGES findings. 

In order to calculate the expected PBH contribution to the radio background I assume that each black hole has a radio emission following the fundamental plane relation between X-ray luminosity, radio luminosity and black hole mass found by \cite{2003MNRAS.345.1057M}. I use the parameterization for radio-quiet AGN from \cite{2006ApJ...645..890W}: log(L$_R$)=0.85 log(L$_X$)+0.12 log(M$_{PBH}$), where L$_R$ is the 1.4 GHz radio luminosity in units of 10$^{40}$ erg/s, L$_X$ is the 0.1--2.4 keV X--ray luminosity in units of 10$^{44}$ erg/s, and M$_{PBH}$ is the PBH mass in units of 10$^8$ M$_\odot$. 
The X--ray luminosity is calculated from the bolometric luminosity shown in figure \ref{fig:Lbol} (right). Assuming the ADAF radiation spectrum above, the fraction of the bolometric luminosity radiated in the 0.1-2.4 keV band is 0.23. For the radio spectrum I assume a power law with spectral index -0.62. This means that the bolometric correction is $\propto$(1+z)$^{0.38}$. The radio luminosities derived this way as a function of PBH mass and redshift are shown in figure \ref{fig:LRadio}. Multiplying these luminosities with the PBH density over cosmic time, converting into observed fluxes and integrating over mass and redshift I obtain a contribution of radio-quiet PBH to the observed radio background of $\sim$3$\times$10$^{-25}$ W m$^{-2}$ sr$^{-1}$ at 1.4 GHz, i.e. a fraction of 0.1\% of the observed synchrotron radio background. Most of this additional radiation field is accumulated at redshifts z$\gtrsim$20. Following \cite{2018ApJ...858L..17F}, this excess radio flux would increase the depth of the 21cm absorption line only by about 30\%. If, however, some fraction of the PBH would be radio-loud (e.g. 5\% with 1000 times higher fluxes), like observed in the AGN population, the 5\% excess high-redshift radio background flux necessary to explain the EDGES feature could be easily achieved by PBH.

\subsection{Primordial Black Holes in the Galactic Center}

Next I discuss some observational effects of the putative PBH population residing in the Galactic Center region. First, assuming a Milky Way dark matter halo of $\sim$10$^{12} M_\odot$, the PBH mass spectrum from section 2 (figure \ref{fig:fPBH}) indeed predicts about one supermassive PBH with a mass $\gtrsim10^{6.5} M_\odot$, consistent with the Sgr A$^*$ SMBH in the center of our Galaxy \citep{2010RvMP...82.3121G}. To estimate the density of dark matter and baryons in the Galactic bulge region itself, I refer to dynamical models of the Milky Way’s center, using the density of red clump giant stars stars measured in infrared photometric surveys, as well kinematic radial velocity measurements of M-giant stars in the Galactic bulge/bar constructed in \cite{2015MNRAS.448..713P}. From N--body simulations of stellar populations for barred spiral discs in different dark matter halos these authors were able to determine with high precision the mass in a volume of ($\pm2.2\times\pm1.4\times\pm1.2$ kpc$^3$) centered on the Galactic Bulge/Bar. The total mass is (1.84$\pm$0.07)$\times$10$^{10}$ M$_\odot$. Depending on the assumed model, about 9--30\% consists of dark matter, i.e. 1.7--5.5$\times$10$^{9}$ M$_\odot$. Applying the above PBH mass spectrum, we thus expect 5--10 intermediate-mass PBH with M$_{PBH}$>10$^4$ M$_\odot$ in the Galactic bulge region, but zero with M$_{PBH}$>10$^5$ M$_\odot$. 

\vskip 0.1 truecm
Recent high-resolution observations of high-velocity compact clouds (HVCC) in the central molecular zone of our Milky Way with the Atacama Large Millimeter/submillimeter Array (ALMA) have indeed identified five promising IMBH candidates, wandering through the patchy ISM in the Galactic Center \citep[see][]{2020ApJ...890..167T}. The most compelling case is HCN–0.044–0.009, which shows two  dense  molecular gas streams in Keplerian orbits around a dark object with a mass M$_{IMBH}$=(3.2$\pm$0.6)$\times$10$^4$ M$_\odot$ \citep{2019ApJ...871L...1T}. The semimajor axis of these Keplerian streams are around 2 and 5$\times$10$^{17}$ cm. Another interesting case is the infrared and radio object IRS13E, a star cluster close to the Galactic Center potentially hosting an IMBH \cite{2005ApJ...625L.111S}. ALMA observations identified a rotating, ionized gas ring around IRS13E \citep{2019PASJ...71..105T}, with an orbit radius of 6$\times$10$^{15}$ cm and a rotation velocity of $\sim$230 km/s. This is thus another promising IMBH candidate with a mass of M$_{IMBH}$=2.4$\times$10$^4$ M$_\odot$.    

Two of the five IMBH candidate sources in \cite{2020ApJ...890..167T} are possibly associated with X--ray sources detected in the deep {\em Chandra} images of the Galactic Center \cite{2009ApJS..181..110M}. IRS13E has the X--ray counterpart CXOGC 174539.7-290029 with an X--ray luminosity L$_{2-10 keV}$$\approx$3$\times$10$^{30}$ erg/s, and CO--0.31+0.11 has the possible X--ray counterpart CXOGC 174426.3-290816 with an X--ray luminosity L$_{2-10keV}$$\approx$4$\times$10$^{29}$ erg/s. The other three sources have X--ray upper limits in the range of several 10$^{30}$ erg/s. Assuming a bolometric correction factor of 1/30 for the 2--10 keV range, the combination of the mass accretion eigenvalue $\lambda$ and the radiative efficiency $\eta$ therefore has to be extremely small, on the order of 3$\times$10$^{-11}$. This is about a factor of 100 lower than the 2$\times$10$^{-9}$ L$_{Edd}$ derived for the Galactic Center black hole Sgr A$^*$ \cite{2014ARA&A..52..529Y}. Even assuming a very low efficiency ADAF model, a steady-state accretion solution is unlikely for these objects. The solution of this puzzle may come from the fact, that the velocity and density gradients of the gas in the Galactic Center region are so large, that the angular momentum forces any accreted matter into Keplerian motion well outside the classical Bondi radius \citep[see][]{2017PhRvD..96h3524P}. Indeed, the orbital periods and lifetimes of the Keplerian streams around HVCCs are in the range 10$^{4-5}$ years, and thus accretion is expected to be highly variable on very long time scales. Another possibility to halt accretion for a considerable time is the feedback created by outflows during efficient accretion events. Numerical simulations of the gas dynamics in the center of the Galaxy \cite{2015MNRAS.450..277C} show that the outflows significantly perturb the gas dynamics near the Bondi radius and thus substantially reduce the capture rate. The net result of both these effects would be a highly variable, low duty cycle bursty accretion onto the IMBH and SMBH in the Galactic Center, consistent with the extremely low accretion efficiencies observed. The accretion limits for black holes in the mass range M$_{PBH}$=20--100 M$_\odot$ derived from deep {\em Chandra} and radio observations of the Galactic Center \cite{2019JCAP...06..026M}, are already shown in figure \ref{fig:fPBH} to be consistent with the assumed PBH mass spectrum. Recent {\em NuSTAR} observations of the Galactic Center, including the effects of gas turbulence  and the uncertainties related to the dark matter density profile even further weaken these constraints \citep{2018A&A...618A.139H}. At any rate, the assumed PBH mass distribution of section 2 is fully consistent with the observational constraints for all PBH masses >20 M$_\odot$ in the Galactic Center. 

\vskip 0.1 truecm
Finally, I check the PBH predictions for lower masses against the Galactic ridge X--ray emission (GRXE), an unresolved X--ray glow at energies above a few keV discovered almost 40 years ago and found to be coincident with the Galactic disk. The GRXE in the 2--10 keV band has a background-corrected surface brightness of (7.1$\pm$0.5)$\times10^{-11}$ erg cm$^{-2}$ s$^{-1}$ deg$^{-2}$ which was largely resolved into discrete sources \citep{2009Natur.458.1142R}, with the brightest source having an X--ray luminosity of about 10$^{32}$ erg s$^{-1}$, and the minimum detectable luminosity around 10$^{30}$ erg s$^{-1}$. The integrated emission has a strong iron line from hot plasma at 6.7 keV, and the authors interpret the X--ray emission as coming from a large population of cataclysmic variables and coronally active stars. Using the mass determination in the Galactic bulge/bar above I find that the average baryon density in this region is in the range 17--22 cm$^{-3}$. However, most of these baryons are locked up in stars. In order to estimate the physical conditions of the gas in the Galactic Bulge/Bar region I follow \cite{2007A&A...467..611F}. According to these authors, there are four phases of the interstellar medium in the Galactic center region: (1) a cold molecular phase in Giant Molecular Clouds with temperatures around 50 K and gas densities n=10$^{3.5-4}$ cm$^{-3}$ covering a volume fraction around 1\%; (2) a warm molecular phase with temperatures around 150 K and gas density n=10$^{2.5}$ cm$^{-3}$, covering a volume fraction of $\sim$10\%; (3) an atomic phase with temperatures around 500-1000 K and density $\sim$10 cm$^{-3}$, covering a volume fraction around 70\%, and (4) ionized gas with temperatures 10$^{4-8}$ K and an unknown density. Depending on the temperature of the interstellar medium, the sound speeds are in the range $c_s$=1--100 km/s. The stellar velocity dispersion in the central region of our Galaxy is in the range 100--140 km/s \citep{2018A&A...616A..83V}, while the dark matter velocity dispersion is about 110 km/s \citep{2002ASPC..276..453K}. In the spirit of the discussion leading up to equation 3.9 and figure \ref{fig:TempVel} (right) above, I assume an effective velocity for Bondi accretion $v_{eff}\approx$50 km/s and $\lambda$=0.1. As shown in figures \ref{fig:Lbol} and \ref{fig:Fbol}, the PBH emissivity for the assumed mass spectrum is typically dominated by objects with M$_{PBH}$>100, which already are discussed above. Indeed, calculating the Bondi accretion rates and radiative efficiencies for objects with M$_{PBH}$<100 for the four ISM phases in the Galactic Center I obtain negligible PBH contributions to the total GRXE brightness. Some individual M$_{PBH}$$\sim$100 M$_\odot$ objects in high density regions could in principle have X--ray luminosities up to $L_{2-10 keV}$=10$^{33}$ erg/s, more luminous than the brightest X--ray source detected in the Galactic Ridge survey \cite{2009Natur.458.1142R}, but taking into account the strong variability and small duty cycle expected for this class of objects, their absence in the surveys is understandable. Some of the fainter unidentified sources in the current deep X--ray surveys could indeed be accreting PBH and future large X--ray observatories like {\em ATHENA} \citep{2013arXiv1306.2307N} or {\em LYNX} \citep{2019SPIE11118E..0KS} should be able to identify more. See also \cite{2019MNRAS.489.2038I} for future searches in the IR and sub-mm region.

\section{Conclusions and Outlook}

The interpretation of cold dark matter as the sum of contributions of different mass PBH families \cite{2019rsta.377..2161G} could explain a number of so far unsolved mysteries, like e.g. the massive seed black holes required to create the supermassive black holes in the earliest QSOs \citep{2007ApJ...665..187L}, the ubiquitous massive LIGO/VIRGO massive binary black holes \citep[e.g.][]{2016ApJ...823L..25K}, or even the putative "Planet X" PBH in our own Solar System \citep{2019arXiv190911090S}. The most abundant family of PBH should be around the Chandrasekhar mass (1.4 M$_{\odot}$). This prediction may already have been vindicated by the recent OGLE/GAIA discovery of a sizeable population of putative black holes in the mass range 1-10 M$_{\odot}$ \citep{2019arXiv190407789W}. Here I estimate the contribution of baryon accretion onto the overall PBH population to various cosmic background radiations, concentrating first on the crosscorrelation signal between the CXB and the CIB fluctuations discovered in deep {\em Chandra} and {\em Spitzer} surveys \cite{2013ApJ...769...68C}. Assuming Bondi capture and advection dominsted disk accretion with reasonable parameters like baryon density and the effective relative velocity between baryons and PBH over cosmic time, as well as appropriate accretion and radiation efficiencies, I indeed predict a contribution of PBH consistent with the residual X--ray fluctuation signal. This signal peaks at redshifts z$\approx$17--30. The PBH contribution to the 2--5 $\mu$m CIB fluctuations, however, is only about 1\%, so that these would have to come from star formation processes associated with the PBH.

\vskip 0.1 truecm
I discuss a number of other phenomena, which could be significantly affected by the PBH accretion. Magnetic
fields are an essential ingredient in the Bondi accretion process, and I argue that the PBH can play an
important role in amplifying magnetic seed fields in the early universe and maintaining them until the
galactic dynamo processes set in. Next I study the contribution of the assumed PBH population to the
re-ionization history of the universe and find that they do not conflict with the stringent ionization
limits set by the most recent {\em Planck} measurements \citep{2018arXiv180706209P}. X--ray heating from the
PBH population can provide a contribution to the entropy floor observed in groups of galaxies
\cite{1999Natur.397..135P}. The tantalizing redshifted 21-cm absorption line feature observed by   
{\em EDGES} \citep{2018Natur.555...67B} could well be connected to the radio emission contributed by PBH
to the cosmic background radiation.
Finally, the number of IMBH and the diffuse X--ray emission in the Galactic Center region are not violated by the PBH dark matter,
on the contrary, some of the discrete sources in the resolved GRXE could be accreting PBH.

\vskip 0.1 truecm
It is obvious, that our simple PBH toy model for the dark matter requires significantly more work to turn it into quantitative predictions. Real magnetohydrodynamic simulations of the whole PBH mass spectrum including their own hierarchical clustering would be required to obtain the full history of their contribution to the cosmic backgrounds. The exciting {\em EDGES} discovery definitely requires a full-blown analysis of the radio contribution of PBH to the cosmic background. Future X--ray observations with {\em eROSITA} and {\em ATHENA}, infrared wide field surveys with {\em Euclid} and {\em WFIRST}, and microlensing observations with {\em WFIRST} will provide important additional diagnostics in this exciting and dramatically developing PBH field (see \cite{2019BAAS...51c..51K,2019ApJ...871L...6K}).  

\vskip 0.2 truecm
\acknowledgments

I am thankful to Juan Garc{\'\i}a-Bellido for sharing a digital copy of the new running spectral index PBH mass distribution model in figure \ref{fig:fPBH} in advance of publication, as well as many very useful discussions about PBH. I am indebted to Matthias Bartelmann for computing the small-scale non-linear relative velocity dispersion (figure \ref{fig:TempVel} right) and providing very valuable comments and corrections to the manuscript. I would like to thank Sergey Karpov for very helpful discussions and inputs about their spherical accretion model. I would also like to thank my colleagues Nico Cappelluti, Sasha Kashlinsky and Alexander Knebe for very helpful discussions and contributions. Finally, I thank an anonymous referee for pointing out a substantial flaw in the first version of the paper, which has been corrected here and led to significant improvements. Throughout this work I made use of Ned Wright's cosmology calculator \citep{2006PASP..118.1711W} and the NASA Astrophysics Data System (ADS), operated by the Smithsonian Astrophysical Observatory under NASA Cooperative Agreement NNX16AC86A.

\bibliographystyle{JHEP} 
\bibliography{PBHXRB_rev.bib} 

\end{document}